\documentclass[floatfix,%
 reprint,
 amsmath,amssymb,
 aps,
]{revtex4-2}

\usepackage{graphicx}
\usepackage{dcolumn}
\usepackage[colorinlistoftodos]{todonotes}
\usepackage{bm}
\usepackage{ulem}
\usepackage{xcolor} 

\usepackage{tabularx}
\usepackage{multirow}
\usepackage{appendix}
\usepackage{array}

\begin{document}

\preprint{APS/123-QED}

\title{Insights from Educators on Building a More Cohesive Quantum Information Science and Engineering Education Ecosystem}

\author{Shams El-Adawy\textsuperscript{1,2}}

\author{A.R. Pi\~na\textsuperscript{3}}

\author{Benjamin M. Zwickl\textsuperscript{3}}

\author{H. J. Lewandowski\textsuperscript{1,2}}

\affiliation{\textsuperscript{1}JILA, National Institute of Standards and Technology and the University of Colorado, Boulder, Colorado 80309, USA}
\affiliation{\textsuperscript{2}Department of Physics, University of Colorado, Boulder, Colorado 80309, USA }
\affiliation{\textsuperscript{3}School of Physics and Astronomy, Rochester Institute of Technology, Rochester, NY 14623, USA}

\date{\today}

\begin{abstract}
As the need for a quantum-ready workforce grows, educators in Quantum Information Science and Engineering (QISE) face the challenge of aligning their programs and courses with industry needs. Through a series of interviews with program directors and faculty across 15 different institutions, we identified the considerations that educators are currently addressing as they develop their various courses and programs. Grounded in a curriculum framework, we conducted a Strengths, Weaknesses, Opportunities and Threats (SWOT) analysis, which revealed shared challenges and opportunities about program context, curriculum development, collaboration, program data collection and evaluation, and connections across stakeholders in the quantum ecosystem that educators should consider when developing their QISE efforts. Our findings highlight five overreaching themes: (1) the strategic ways educators navigate institutional structures to support QISE initiatives, (2) the ongoing challenge of aligning QISE curricula with industry and institutional needs, (3) the importance of fostering interdisciplinary collaboration across departments and institutions in QISE, (4) the need for robust data collection and evaluation to inform QISE course and program development, and (5) the importance of strengthening industry-academia connections to prepare students for the quantum workforce. The details and interconnections in our findings illustrate the value of applying a structured approach to QISE course and program development with the goal of creating a more cohesive QISE education ecosystem.
\end{abstract}

\maketitle


\section{Introduction}

Quantum Information Science and Engineering (QISE) is a field forged in physics, spanning across disciplines. QISE lays the foundation for emerging quantum technologies, such as quantum sensing, quantum networking and communication, and quantum computing. In recent years, QISE research and education have received significant attention globally \cite{NQI,riedel2019europe, knight2019uk}. 

In the United States, several landmark pieces of federal legislation have positioned QISE at the forefront of strategic investment. First, the National Quantum Initiative (NQI) Act established quantum technology as a federal priority and advocated for the support and development of quantum technology research \cite{NQI}. Second, the Quantum Information Science and Technology Workforce Development National Strategic Plan highlighted the need for higher education to support the development of the quantum workforce through the expansion of QISE courses and programs \cite{NSP}. Third, the  CHIPS and Science Act underlined the need to assess the landscape of QISE education to support the implementation of the strategic plans \cite{CHIPS,taylor2023us}. 

As a result of this significant investment, various stakeholders in the quantum education ecosystem (curriculum developers, faculty members, and program leaders) are developing courses, minors, and master's programs aimed at preparing students for the emerging quantum workforce \cite{perron2021quantum,economou2022hello, asfaw2022building, blanchette2024design, goorney2024quantum, qerimi2025exploring}. Educators and education researchers have focused on creating QISE specific-curricula \cite{devore2016development, economou2020teaching, salehi2021computer}, characterizing the QISE education  landscape \cite{cervantes2021overview, buzzell2025quantum, Pina2025}, and understanding the needs of the quantum industry to prepare students to enter this workforce \cite{fox2020preparing, aiello2021achieving, kaur2022defining, hughes2022assessing, hasanovic2022quantum}.

Although these various initiatives in QISE education and workforce development are valuable, they are not as connected as they could be, which challenges the community's ability to structure and build a cohesive QISE education ecosystem.  For educators who are interested in creating a QISE course or program in their institutions, few structured frameworks or approaches exist to support them in designing and implementing  a QISE course or program. Experienced QISE educators may benefit from understanding the experiences of others directly engaged in the implementation of QISE courses and programs. By placing educators at the center of this research, we aim to showcase how they navigate the evolving QISE education ecosystem and offer a structured approach to developing the field. 

Our goal with this paper is to demonstrate the use of a structured framework for designing and implementing QISE programs and courses, which will allow the community to learn from other educators and strategically address gaps and challenges in the field. By integrating a structured curriculum framework with a Strengths, Weaknesses, Opportunities and Threats (SWOT) analysis, we highlight key factors that educators are considering in their own QISE efforts. In particular, we address the following two research questions: \begin{itemize}
    \item How are educators designing, implementing, and sustaining QISE programs and courses in an effort to better support students’ education?
    \item What insights do educators' experiences offer regarding the future development of QISE education?
\end{itemize}

Recognizing that many QISE efforts are still emerging, our goal is to offer an overview of the key affordances and challenges educators face --- both to support physics faculty getting started in QISE education and to offer a broad synthesis for those who have been engaged in this field for some time. Our work does not aim to evaluate specific QISE courses or programs, nor does it advocate for social or cultural changes within existing initiatives.

This paper begins by situating our work within the QISE education research and curriculum development literature in Sec. \ref{Background}. Then, within the same section, we explain the reasoning behind our use  of framework and methodology to answer our research questions. Next, we detail the process of data collection and analysis in Sec. \ref{Methods}. In Sec. \ref{results}, we structure the findings around each component of the framework, first for QISE course development and then QISE program development. Lastly, in Sec.\ref{discussion}, we discuss the key themes and implications for the QISE education community.  

\section{Background}\label{Background}
\subsection{Research in QISE education}
The rapid advancement in QISE has motivated significant research in QISE education, primarily driven by the need to foster interest and prepare students for the quantum workforce. Workforce development studies have consistently highlighted the need for universities to align curricula with the skills and degrees needed in the rapidly evolving quantum workforce \cite{fox2020preparing, aiello2021achieving, kaur2022defining, hughes2022assessing, hasanovic2022quantum, greinert2023future, greinert2023towards, greinert2024advancing}. Some of these studies focus on how traditional academic pathways, such as PhD programs in physics, already provide an entry point into the quantum industry \cite{fox2020preparing, aiello2021achieving}. However, industry stakeholders emphasize an increased demand for bachelor's degree graduates to enter the quantum industry, and increased quantum awareness and workforce readiness across educational levels \cite{fox2020preparing, aiello2021achieving}. This growing emphasis on quantum workforce  preparedness at all academic levels has motivated deeper research into how educational programs can better equip students with relevant skills. 

As a result, a growing wave of research focuses on the design and delivery of QISE courses, identifying both opportunities and challenges in QISE curriculum development.  Studies have identified core topics---including superposition, entanglement, quantum algorithms and quantum communication---that could serve as foundational components for a standardized introductory QISE curriculum \cite{meyer2024introductory, meyer2022today}, which suggests a pathway toward greater coherence in QISE education. Furthermore, efforts are emerging to coordinate curriculum development across institutions, emphasizing the importance of modular and adaptable course structures and content to support educators \cite{barnes2025outcomes}. These efforts aim not only to standardize foundational QISE content, but also to strengthen the overall coherence and accessibility of QISE education.

Despite this momentum, significant challenges persist in QISE course development. One major concern is the varied preparedness levels in QISE courses of students coming from a range of disciplines. Balancing accessibility for students with limited quantum background, while maintaining academic rigor poses an ongoing tension for instructors \cite{meyer2022today}. Additionally, given that QISE is inherently interdisciplinary, many faculty members do not have expertise across all QISE domains, making it difficult to balance breadth and depth in course content \cite{meyer2022today}. Research suggests that a more structured and systematic approach to content development could help mitigate these challenges and lower barriers to course creation \cite{goorney2024framework}. 

Beyond individual QISE courses, researchers have also examined the broader development of QISE programs, revealing disparities in who has access to formal QISE education. QISE degree programs are disproportionately concentrated in PhD-
granting schools \cite{aiello2021achieving, meyer2024disparities, Pina2025}. Recent landscape studies of QISE education show that interdisciplinary programs are the most common ones present in the US \cite{Pina2025, cervantes2021overview}. 

In an effort to expand accessibility, researchers are proposing various structural approaches to support QISE program development. 
One approach suggests embedding QISE content within existing STEM programs. For instance, Asfaw \textit{et al.} \cite{asfaw2022building} advocate for integrating QISE topics within existing STEM curricula, such as through minors or specialized tracks, to create a more inclusive and accessible model for QISE education. This strategy allows institutions to introduce QISE content without requiring the extensive resources needed to establish standalone programs.
Alternatively, some researchers propose designing dedicated undergraduate QISE degrees from the ground up. Blanchette \textit{et al.} emphasize the importance of a data-driven approach to curriculum design in developing full-fledged QISE undergraduate programs, ensuring that students gain the skills necessary for careers in quantum technologies \cite{blanchette2024design}. While these approaches mark important first steps towards structuring QISE education and building QISE capacity, current research has not deeply investigated the experiences of educators responsible for implementing these programs, particularly the constraints and decision-making processes they navigate in developing QISE programs. 

These existing studies have made important strides in understanding workforce needs, curriculum development, and programmatic structures. We build on this past research by focusing on an under-explored piece, which is the experiences of instructors and program directors who are actively shaping these programs. Understanding these experiences and how they navigate them is essential for building a more cohesive and sustainable QISE education ecosystem --- one that not only reflects industry needs, but also supports the educators tasked with translating those needs into meaningful learning experiences through courses and programs.

\subsection{Research in curriculum and program development}

To support the structured development of QISE courses and programs, we situate our work within curricula and program development in physics education research (PER) and Science Technology, Engineering and Mathematics (STEM). We draw on key frameworks that inform curriculum and program development, which fall into three categories: instructional design for course transformation, organizational and culture change of a department or institution, and program assessment. As we will discuss in this section, while these frameworks provide valuable insights, they operate independently, at different structural levels (course level, department-level, or institution-level) and with distinct goals (make research-based course changes, facilitate change around a long-standing departmental or institutional problem, or maintain accreditation standards), which makes it challenging to use them to analyze both QISE courses and programs in a cohesive and integrated ecosystem. 

In the domain of instructional design, several models exist to guide course development. One of the most widely used frameworks is ADDIE, which stands for Analyze, Design, Develop, Implement and Evaluate \cite{ADDIE}. ADDIE provides a structured and iterative approach to conceptualize and refine course material \cite{ADDIE}. Moreover, research in physics education has led to models specifically focused on science education, such as the Science Education Initiative (SEI)  model \cite{chasteen2011thoughtful, chasteen2015educational}. SEI emphasizes a research-driven approach to course transformation by defining learning goals, mapping assessment outcomes and identifying instructional approaches \cite{chasteen2011thoughtful}. Although ADDIE and SEI frameworks offer valuable methodologies for course design, their primary focus is on helping instructors develop and improve individual courses. These frameworks are not designed to investigate how educators develop both courses and programs at a broader scale. Given that our goal is to provide a broad synthesis of key opportunities and challenges QISE educators face, the scope and intent of frameworks such as ADDIE and SEI do not fully align with our research objectives, as they are centered solely on course development. 
 
Organizational and culture change frameworks provide insights into educational transformation. One such model is the Four Frames for Systemic Change in STEM departments, which focuses on institutional change \cite{reinholz2018four}. The framework posits that effective sustainable change in a department requires integrating four perspectives: structural policies, human resources, political dynamics, and cultural facets. Another relevant framework is the Collective Impact Framework \cite{Kania2011}, which provides a structured approach to solving complex social problems by aligning stakeholders towards a unified goal. However, while these frameworks are insightful for guiding institutional transformation, our work does not seek to drive social or cultural change within institutions, as we do not have data from multiple stakeholders within higher education to be able to appropriately address this topic. Instead, we focus on understanding the experiences of educators as they navigate the opportunities and challenges of QISE course and program development.
 
Program assessment frameworks also play a critical role in shaping curriculum development. For example, a common approach is the ABET accreditation process, which ensures that academic programs meet established standards \cite{ABETAccreditation}. While ABET offers a valuable framework for evaluating existing programs, it does not provide ways to understand the unique challenges educators face when designing new interdisciplinary programs, particularly in an emerging field such as QISE. ABET focuses on ensuring compliance with predefined criteria rather than exploring the considerations that come up during course or program development. Another example is the Higher Learning Commission (HLC), which focuses on assessing whether institutions' overall mission, governance, and resources meet established standards \cite{HLC}. However, like ABET, HLC centers primarily on meeting standards for institutional accreditation rather than understanding the challenges and opportunities of course or program development.
 
Taken together,  these frameworks provide important tools for course transformation, institutional change, or program assessment. However, none fully address both course and program levels structures to consider when designing and implementing QISE content and structures. Thus, we turn to broader literature to identify a framework that allows us to analyze and contextualize the unique challenges and affordances of QISE educators' experiences.  

\subsection{Framework}\label{framework}

We turn to a systems-thinking approach to curriculum design, which has been used in health professional education \cite{khanna2021designing}. Specifically, we adapt  the \textit{3P-6Cs  framework} developed by Khanna \textit{et al.} \cite{khanna2021designing}, which conceptualizes an educational program as a complex adaptive system through which students navigate to become part of a community of professionals. 

We have adapted this framework for two main reasons. First, as a relatively new and rapidly evolving field, QISE lacks structured ways to address the many factors shaping education in the field. A structured approach helps us understand the various factors that come into play in the development process. Second, many of the recent QISE efforts are driven by perceptions of the workforce needs for the quantum industry. Similarly to the health care profession, where academic preparation focuses on developing the knowledge and skills needed to become a medical professional, much of the recent academic preparation in QISE is increasingly driven by the need to prepare students to become professionals in the quantum industry.  

\begin{figure}[t]
    \centering
    \includegraphics[width=\columnwidth]{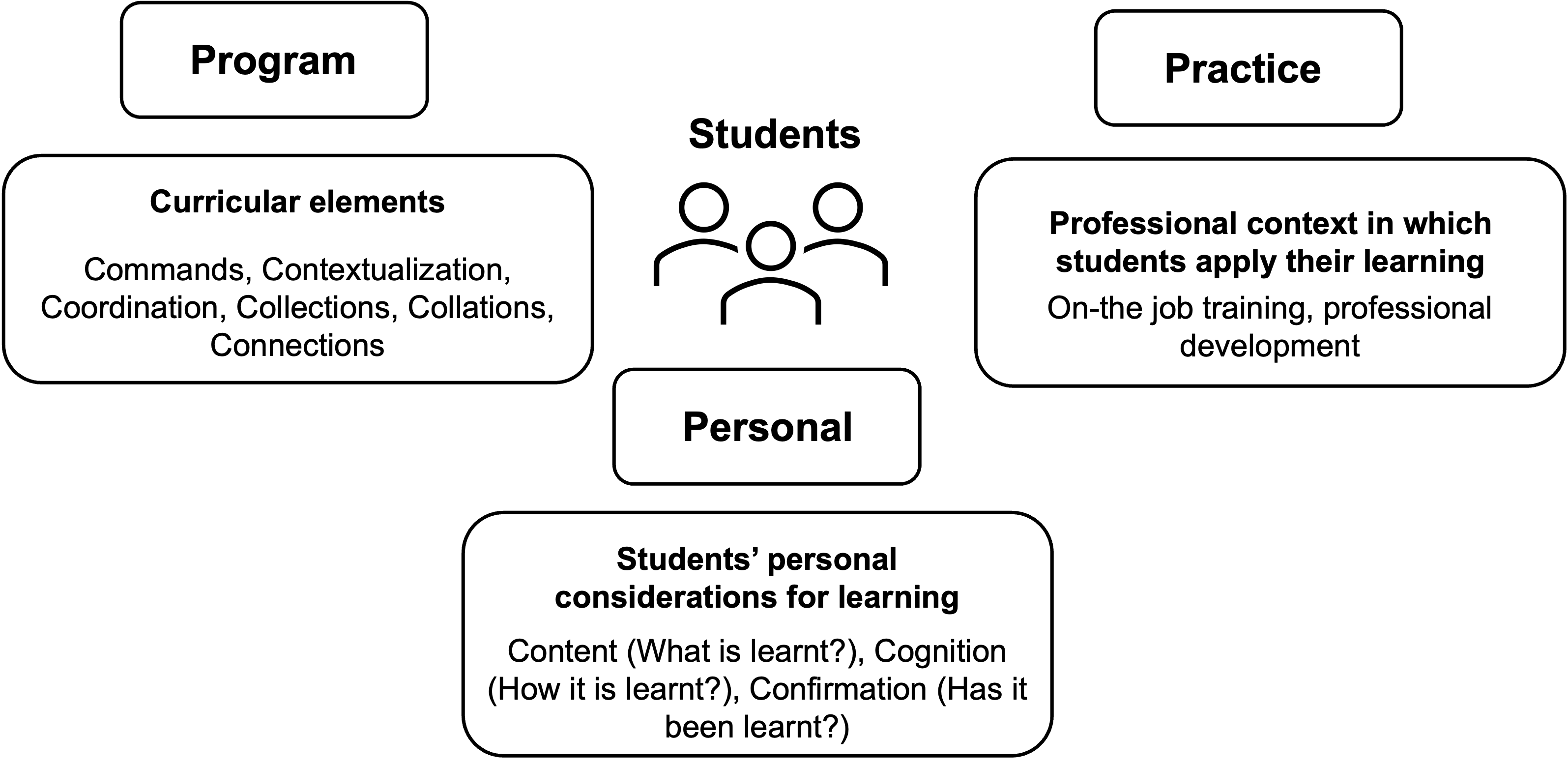}
    \caption{Illustration of the \textit{3P-6Cs framework} \cite{khanna2021designing}, highlighting how educational programs can be structured around three interconnected level (Program, Personal, and Practice) that students navigate to become professionals in their field.}
    \label{fig:illustrationframework}
\end{figure}

As illustrated in Fig. \ref{fig:illustrationframework}, the \textit{3P-6Cs framework} allows us to organize the key factors educators need to address to support students' education. The framework structures educational programs into three interconnected levels:
\begin{itemize}
    \item Personal (P1): The individual learner's experiences and background in navigating the curriculum.
    \item Program (P2 with 6Cs): The explicit and implicit curricular elements that shape a learner's educational journey.
    \item Practice (P3): The broader professional context in which students apply their learning.
\end{itemize}

Our study focuses specifically on the Program level. While this structured framework encourages holistic perspectives, it also highlights that complex systems are best understood by examining their components. By starting with the program aspect, we are consistent with the framework’s intent, while laying the groundwork for future research that will focus on the personal and practice components. For example, future work from our team and other researchers will examine students' learning experiences in QISE courses and quantum industry professionals' experiences in the workforce.

In operationalizing the components of the program level of the \textit{3P-6Cs framework}, we defined each component to meet the specific needs and objectives of quantum education, which involved translating the framework’s general components into specific practices tailored to QISE education. As indicated in the name of the framework, the original framework components start with the letter C. \textit{Commands} refers to the clear vision and mission of an educational offering, which we simplified to \textit{goals}. In our context, \textit{goals} pertain to the vision, mission, and objectives of a QISE course or program. \textit{Contextualization} focuses on curriculum development, which we refer to simply as \textit{curriculum} tailored to institutional resources and needs in QISE. \textit{Coordination} refers to collaboration among QISE educators, which we refer to simply as \textit{Collaborations}. \textit{Collections} and \textit{collations} pertain to data collection and evaluation to ensure QISE courses and programs are meeting their goals. For simplicity, we labeled them  \textit{collection} and \textit{evaluation}. Finally, \textit{connections} involve partnerships across the quantum ecosystem (e.g., institutions and quantum companies). Our operational definitions are summarized in Table \ref{tab:framework_definitions}. 

While this curriculum framework provides a structured approach to identifying QISE course and program design, it does not inherently capture the challenges and opportunities that arise during the development process. To address this, we combine the framework with a SWOT (Strengths, Weaknesses, Opportunities and Threats) analysis \cite{sammut2015swot}, which allows us to systematically identify obstacles and opportunities in QISE course and program development.

\subsection{ Strengths, Weaknesses, Opportunities, and Threats (SWOT) analysis}
SWOT is a methodology that supports strategic planning and decision-making about program structure and organization \cite{sammut2015swot}. In STEM education, SWOT has been used to improve teaching and learning, particularly by guiding program development and improvement \cite{orr2013conducting}. In our context, it helps pinpoint internal and external factors that impact QISE programs and courses, allowing us to address our research questions. In particular, we broadly define each element of SWOT as follows:
\begin{itemize}
    \item Strengths: Internal attributes, structures, or processes within an institution that positively contribute to the effective development and sustainability of a QISE course or program.
    \item Weaknesses: Internal limitations or structural barriers within an institution that may hinder the design, execution, or refinement of a QISE course or program.
    \item Opportunities: External trends or developments that present potential for enhancement or strategic alignment in support of the development of a QISE course or program.
    \item Threats: External challenges or disruptions that may negatively impact the sustainability or effectiveness of a QISE course or program. 
\end{itemize}
 To ensure clarity and consistency in our analysis, we developed detailed operational definitions of SWOT that apply to each component of the curriculum framework. These definitions are provided in Appendix \ref{appendix:swot_definitions}.  

\begin{table*}
\caption{Components of the \textit{3P-6Cs framework}, the abbreviated definitions associated with each component (complete definitions can be found in \cite{khanna2021designing}) and the operationalized definitions within the context of QISE.}
\begin{tabular}{p{8.5cm} @{\hspace{0.5cm}} p{8.5cm}}
\toprule
\multicolumn{2}{c}{} \\[-0.5em]
\textbf{Components and definitions in medical education} & \textbf{Operationalized components and definitions in QISE} \\
\hline

\textbf{Commands}: High-level vision, mission, outcomes, and guiding practices for courses and programs 
& \textbf{Goals}: Overarching vision, mission, and objectives of QISE courses and programs \\
[1em]

\textbf{Contextualization}: Curricular themes, content, and pedagogical approaches shaped by contextual factors 
& \textbf{Curriculum}: Customizing the QISE curriculum to align with institutional resources, priorities, and specific needs \\
[1em]

\textbf{Coordination}: Collaborative efforts to integrate learning, teaching, and educational activities 
& \textbf{Collaboration}: Coordinating efforts among QISE educators within and across institutions to align curriculum design and instruction practices \\
[1em]

\textbf{Collections}: Systematic gathering of data such as students' performance and retention metrics 
& \textbf{Collection}: Collecting data on students' academic performance and career preparation in QISE courses and programs \\
[1em]

\textbf{Collations}: Synthesizing collected data to inform decision making and guide course and program improvement 
& \textbf{Evaluation}: Synthesizing collected data to provide comprehensive assessment of student learning in QISE courses and QISE programs' effectiveness \\
[1em]

\textbf{Connections}: Fostering network among stakeholders to support students for professional practice 
& \textbf{Connections}: Building partnerships between academia and the broader quantum ecosystem, including industry and government stakeholders, to ensure curricula relevance and workforce preparedness \\
\toprule
\end{tabular}
\label{tab:framework_definitions}
\end{table*}

By integrating a SWOT analysis with the curriculum framework, we ensure that identified SWOT factors are not viewed in isolation, but as interconnected components within a broader educational ecosystem. This approach provides a community-level perspective on designing and sustaining effective QISE courses and programs. Prior research highlights that combining a SWOT analysis with a curriculum framework enhances the understanding of overarching themes and interdependencies, enabling programs to adapt to the evolving landscape of higher education \cite{awuzie2021facilitating, allareddy2024orthodontic}. Hence, conducting the SWOT analysis at this community level allows us to identify shared challenges and opportunities across different courses and programs. In turn, it provides us with insights into overarching patterns that may not be visible within an individual program or course.  

\section{Methods} \label{Methods}
\subsection{Data Collection}
\subsubsection{Recruitment}
To identify educators engaged in QISE to participate in our study, we used a combination of convenience sampling and purposeful sampling \cite{etikan2016comparison}. Leveraging our network and experience in the QISE field, we identified QISE educators based in the United States who had participated in national conferences and workshops focused on quantum education and workforce development, engaged in QISE curriculum development,  and/or had published research in QISE education. Once we gathered a list of about 25 individuals, we reached out to invite them to participate in one-on-one semi-structured interviews for our research.

Fifteen educators agreed to participate. A detailed breakdown of our interview participants can be found in Appendix \ref{appendix:interview_participants}. In brief, their efforts in QISE spanned a wide range of activities: some developed traditional quantum courses with integrated QISE topics, others created and taught standalone QISE courses, while some developed QISE minors or concentrations at the undergraduate or graduate level. Additionally, a few designed and managed standalone master's programs in QISE. Of the 15 participants, 12 were affiliated primarily with physics departments, 2 with engineering, and 1 with biology. According to the Carnegie Classification, 10 participants were from R1 institutions, 2 from R2 institutions, 2 from Doctoral/Professional Universities, and 1 from a Master's College and University. 

\textit{Note on terminology:} Based on our data and in alignment with our previous work \cite{Pina2025}, we define programs ``to include associate's, bachelor's, master's, and PhD degrees and certificates, minors, concentrations, focuses, and tracks within majors'' \cite{Pina2025}. Our interview participants were developing or leading many of these types of programs, including  QISE minors or concentrations at the undergraduate or graduate level, and stand-alone master’s programs in QISE. Similarly, our interview participants were developing or teaching traditional quantum courses with added QISE topics and/or standalone QISE courses.

\subsubsection{Data collection tool and process}
The semi-structured interview protocol was designed to understand how faculty are using data to make curricular decisions. In particular, these interviews were conducted to complement concurrent work we are conducting, where we are assessing the landscape of quantum and QISE education in the United States using publicly available data \cite{Pina2025}. Additionally, these interviews were also intended to guide data collection for the next phase of our project, which aims to assess the QISE industry landscape. Our interview protocol covered the following topics:  (a) the interviewee’s current role and relationship with QISE and details about current/future programs and/or courses in QISE at their institution, (b) the interviewee's familiarity with, and questions about, the quantum industry workforce, and (c)  the interviewee's familiarity with,  and questions about, higher education's QISE landscape. Parts (a) and (c) of the protocols contained the most information pertaining to the analysis for this paper. The list of interview protocol questions can be found in the Supplemental Materials.

Interviews were conducted by the first author over video conference (Zoom), recorded, and transcribed (otter.ai) for analysis. The length of the interviews varied between 35 to 60 minutes depending on how much detail the interviewee gave in their answers.

\subsection{Data Analysis}

\begin{figure*}[t]
    \centering
    \includegraphics[width=\textwidth]{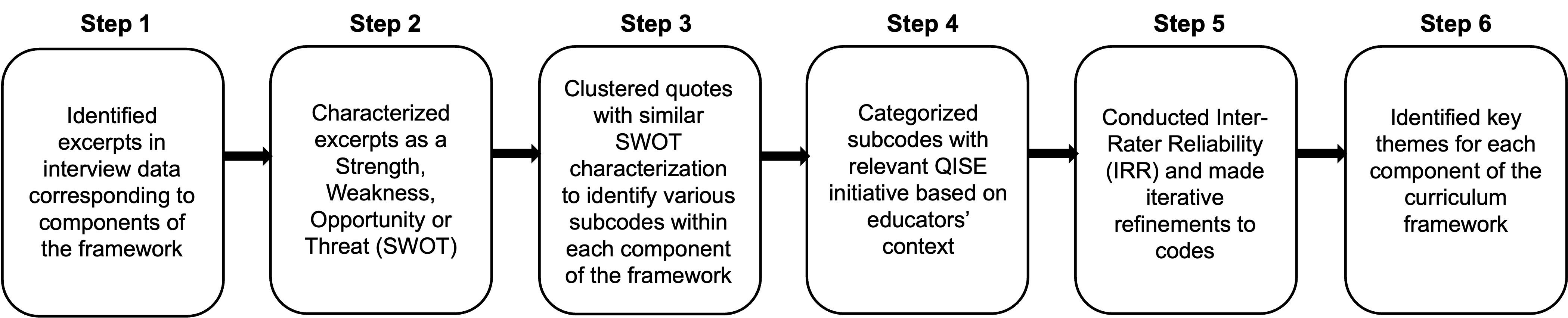}
    \caption{An overview of the different steps of the analysis process from framework application and SWOT analysis to community-level themes of shared challenges and opportunities in the design and development of QISE courses and programs}
    \label{fig:analysissteps}
\end{figure*}
An overview of our analysis process that connects our methods and the two frameworks that led to identifying the key themes encapsulating educators' experiences navigating the QISE education ecosystem is illustrated in Fig. \ref{fig:analysissteps}.

\textbf{Step 1}: The operationalized \textit{3P-6Cs framework} provided a set of \textit{a priori} codes to apply to our interview data. These \textit{a priori} codes are the operationalized components and definitions in QISE listed in Table \ref{tab:framework_definitions}. We read each interview transcript extracting participants’ answers that fell within each of the main components of the framework. From the entire interview corpus, we extracted 247 excerpts that directly pertained to components of the framework. Excerpts were not mutually exclusive, which means that a single excerpt could be categorized under more than one component of the framework. 

\textbf{Step 2}: We applied the a priori codes from SWOT to each of the excerpts categorized in Step 1. The detailed operationalized definitions of SWOT within the framework are listed in Appendix \ref{appendix:swot_definitions}.

\textbf{Step 3} We intersected the codes with similar SWOT characterization within the framework. This grouping allowed us to perform emergent coding, which involved characterizing the various subtleties as sub-codes within each component of the framework and SWOT. For example, the following quote was part of the \textit{Goals} component of the framework, was tagged as a \textit{Weakness} and succinctly described as \textit{Administrative challenges on setting up a program}: \begin{quote}
    \textit{Another hurdle is, if you create a new degree, the question is, do you have the resources? And how is that degree administered? Is there a separate admissions committee advising of the students? Something that doesn't already take advantage of existing apparatus in a department which administers degree programs, if you wanted to create a new degree that straddled existing structures. That's a big hurdle.}
\end{quote}

\textbf{Step 4}: To contextualize the sub-codes identified in Step 3 (all sub-codes are listed in Table \ref{tab:results}), we categorized each one according to the type of QISE initiative (course or program), its stage of development (design or implementation), and the presence of any QISE specific resources at the institution (quantum center, hub, institute).

\textbf{Step 5}: After completing steps 1-4, a round of Inter-Rater Reliability (IRR) was conducted with a second researcher. On 10\% of the data coded with the \textit{3PC-6Cs framework} and SWOT components, there was 70\% observed agreement pre-discussion between the two raters. Post-discussion on the same 10\% of the data, there was 98\% of agreement. The sources of disagreement  were related to two aspects: the meaning of codes within the SWOT and framework, as well as the possibility for excerpts to be double-coded. This was addressed by clarifying that double-coding excerpts is allowed and refining definitions of codes. Observed agreement is sufficient because the focus of our paper is more on existence than on prevalence claims (though we report prevalence for some items to provide deeper analysis). The high observed agreement post-discussion demonstrates strong consistency between raters regarding the presence of codes. 

\textbf{Step 6}: Lastly, we identified themes for each component of the framework across QISE courses and programs. 

The first author examined the data following this process that we just described and summarized in Fig. \ref{fig:analysissteps}. At each step of the process, research team members discussed emerging patterns and ideas identified during the synthesis, making iterative refinements to the articulation of the overarching themes.

\subsection{Limitations}
There are limitations in our data that constrain the scope of the claims and inferences we can make in this paper. First, the lack of varied institutional types in our data limits the scope of interpretation of our community-level claims. Most of the interviewees in our dataset are at R1s. Key perspectives from other types of institutions are needed to make sure we capture all the factors educators at various institutions are facing. However, this limitation reflects a larger trend in QISE education, where most efforts are still largely concentrated at R1s \cite{Pina2025}. Despite this, our dataset includes some perspectives from non-R1 institutions, which provide at least a partial view of the larger landscape.

Another limitation is the lack of non-physics disciplines in our data. Due to our own physics backgrounds and professional networks, most of our interview participants were affiliated with physics departments. Although most QISE efforts remain concentrated in physics departments, QISE is increasingly interdisciplinary \cite{Pina2025}. In particular, there are a growing number of courses and programs in engineering and computer science that we may have not captured in our analysis \cite{pina2025asee, Pina2025}. Although the claims in this paper are primarily situated within the context of physics, they hold broader relevance for QISE education as a whole, given the substantial role that physics plays in shaping and advancing QISE education. 

In addition to these data limitations, our application of the SWOT methodology differs from the conventional approach. While SWOT analyses are typically conducted within the context of an individual course or program, our dataset spans multiple institutions, which makes it difficult to draw conclusions specific to any one course or program. As a result, we applied the SWOT analysis  at the community level, which provides valuable insights into answering our research questions. In fact, each educator in our study contributed insights into at least one strength, one weakness, one opportunity, and one threat, ensuring that our findings reflect a broad range of experiences across various institutional contexts and QISE initiatives. 

\section{Results}\label{results}

We begin by presenting the factors identified for QISE course development, followed by those for QISE programs. These factors are anchored in the \textit{3P-6C} curriculum framework with the SWOT analysis serving as a lens to interpret their significance. This approach showcases the interconnected nature of the strengths, weaknesses, opportunities, and threats highlighted by educators during the interviews.  

Given the scope of our dataset, the factors identified are not universally applicable to every QISE program or course. Instead, they are intended to illustrate the broad range of factors highlighted by our interview participants. Table \ref{tab:results} provides a summary of the key factors discussed in this section.

\begin{table*}
\caption{Key factors that emerged within the curriculum framework and SWOT analysis for educators developing QISE courses and programs}
\begin{tabular}{p{2.2cm} p{8.1cm} p{8.1cm}}
\toprule
\multicolumn{3}{c}{} \\[-0.5em]
\textbf{Framework component} & \textbf{QISE course} & \textbf{QISE program} \\
\hline
\textbf{Goals} & Weaknesses: \begin{itemize}
    \item Administrative hurdles on setting up a QISE course
\end{itemize} & 
Strengths: 
\begin{itemize}
    \item Strategic program development within existing institutional structures
    \item Well thought-out program development
    \item Intentional interdisciplinary program development
\end{itemize}
Weaknesses: 
\begin{itemize}
    \item Administrative hurdles in setting up program
\end{itemize}
Opportunities 
\begin{itemize}
    \item Exploration of quantum workforce needs
    \item Cross-institutional learning
\end{itemize}
Threats
\begin{itemize}
    \item Unclear utility of QISE degree program
    \item Mismatch between academic and industry goals
\end{itemize} \\ \hline
\textbf{Curriculum} & Strengths: \begin{itemize}
    \item Integration of QISE topics into existing quantum courses
    \item Student-centered course design
\end{itemize}
Weaknesses: \begin{itemize}
    \item Curriculum breadth gaps
    \item Varied STEM backgrounds
\end{itemize}
Opportunities: \begin{itemize}
    \item Industry-driven curricula revisions
    \item Sharing QISE content among educators
\end{itemize}
Threats: \begin{itemize}
    \item Industry-academia misalignment on learning objectives
    \item Reliance on proprietary teaching platforms 
\end{itemize} & Strengths: \begin{itemize}
    \item Broad available QISE expertise
\end{itemize}
Weaknesses: \begin{itemize}
    \item Insufficient faculty with QISE expertise
    \item Lab-space constraints
\end{itemize} \\ \hline
\textbf{Collaboration} & Strengths: \begin{itemize}
    \item Continuous professional development
\end{itemize} & Strengths: \begin{itemize}
    \item Cross-departmental faculty collaboration
\end{itemize}
Weaknesses: \begin{itemize}
    \item Departmental silos
\end{itemize} \\ \hline
\textbf{Collection} & & Strengths: \begin{itemize}
    \item Longitudinal data gathering
\end{itemize} 
Weaknesses: \begin{itemize}
    \item Only anecdotal data
\end{itemize} \\ \hline
\textbf{Evaluation} & & Strengths: \begin{itemize}
    \item Summative assessment
\end{itemize}
Opportunities: \begin{itemize}
    \item Evaluation of outcomes from programs
\end{itemize} \\ \hline
\textbf{Connections} & Weaknesses: \begin{itemize}
    \item Limited industry connections
\end{itemize}
Opportunities: \begin{itemize}
    \item Leveraging alumni insights
\end{itemize} & Strengths: \begin{itemize}
    \item Established collaborations with quantum industry
\end{itemize} \\ 
\hline \hline
\end{tabular}
\label{tab:results}
\end{table*}

\subsection{QISE course development}
Educators developing QISE courses in our dataset were taking two main approaches: designing standalone QISE courses or embedding QISE topics into existing quantum courses. Both pathways reflect an effort to define the scope of QISE instruction, while navigating institutional resources and evolving disciplinary boundaries.

Drawing from our curriculum framework, we analyzed QISE course development through the lens of the components that emerged in our analysis, which were Goals, Curriculum, Collaboration, and Connections. Notably, our interview participants did not explicitly reference the Collection or Evaluation components in their discussions of QISE courses. As a result, we included subsections for only four of the six components of the framework. 

\subsubsection{Goals}
The development of QISE courses was shaped significantly by institutional structures. A  \textbf{weakness} highlighted was the \textit{administrative hurdle of setting up a QISE course}. Five of the 15 educators we interviewed described the difficulty of establishing new QISE courses, especially securing regular approvals to offer these courses.  For instance, Casey, a physics faculty at public university, described: 
\begin{quote}
    \textit{We have designed two courses, but we haven't got them approved or regularly offered. One has been offered as a topics course, so kind of a one time shot at linear algebra course was taught kind of as a quantum algorithms type thing. So we are trying to get QIS focused courses approved and in the curriculum. But there is a bunch of bureaucratic challenges.}
\end{quote}
One of the reasons for these bureaucratic challenges, as identified in our data, was the emerging and still evolving status of QISE as a discipline. Without an established curriculum within existing institutional structures, QISE courses faced more challenges in getting approval for regular offering. As a result, we anticipate that institutions may hesitate to commit resources to courses that do not fit seamlessly into existing degree programs. 

\subsubsection{Curriculum}
A key \textbf{strength} in current QISE course development was the \textit{integration of QISE topics into existing quantum courses}. This approach offered a pragmatic entry point for educators because it allowed them to introduce students to QISE concepts without needing to build entirely new courses from scratch. Jules, a physics faculty member at a private university, shared how she fits QISE concepts within her modern physics course:
\begin{quote}
    \textit{So when we start talking about a little bit about quantum mechanics, we talk about the Schrodinger equation, where it came about, talk about infinite wells [...]So talking about an electron in a wire as an approximation for  the infinite well potential, and I think that's where we could then, as an application do a little bit of the principles of quantum computing. So that's the first part of it. And then we build band theory, and we talk about metals and semiconductors and then doping, and by the end, we talk about diodes and transistors. And so that's where the SQUIDs [superconduting quantum interference devices] might come in. So basically as applications.}
\end{quote} 
Five educators in our dataset discussed this strategy of embedding QISE topics into existing quantum courses. This was not an unsurprising choice as QISE topics align well with existing quantum mechanics courses as direct extensions or applications. This strategy aligns with national trends observed in how QISE topics appear in course offerings \cite{Pina2025}. This practice is scalable and accessible, regardless of institution type or available resources, thus can be used by faculty as they get started in QISE. 

Another \textbf{strength} was having a \textit{student-centered course design}  approach. Five educators emphasized the importance of aligning course content with student career aspirations and practical needs. As Casey described: \begin{quote}
    \textit{So I think anytime I'm including stuff in, I am trying to consider how it's going to help the students or benefit them. As much as I like it, it's not for me and I think pedagogically, especially at our institution, where I think just over three quarters of our graduates go straight to work. They're not going to grad school. They want a job, and they're going out to do that. I think highlighting how this is more than like an equation in a book, in just a class, you can say that you need to get that degree, that you can learn some practical things. I think it's better for retention,[...] you have to be aware of where the students are when you're going to try and do these things.}
\end{quote}
This student-centered focus reinforced the importance of collaboratively designing courses that would be most relevant to students.

However, two \textbf{weaknesses} in curriculum development at the course level emerged. The first was \textit{curricular breadth gaps}. Some institutions have QISE courses that do not cover key areas such as quantum sensing. One of our educators, Jaime, who is a physics faculty at public research university and the director of their quantum center, commented on this issue saying: \begin{quote}
    \textit{At the sort of [QISE] literacy course level, we are trying to touch on communications and cryptography, which I think a lot of them have, but also sensing, and I think that's the one thing that I see often getting dropped. And in various discussions, there's been a lot of discussion about: how do you do that?}
\end{quote}
Four educators in our dataset discussed this challenge, which reflects the broader difficulty of developing a well-rounded QISE curriculum. In particular, some research has highlighted how quantum sensing curricula is particularly under-developed \cite{Zwickl2024QuantumSensing, Namitha2025}.

The second \textbf{weakness} was developing QISE courses for students with \textit{varied STEM backgrounds}. Sam, an engineering faculty at a private research university, described this challenge:
\begin{quote}
    \textit{The challenging thing with this class every year is that I'm trying to teach a class for people with no background in quantum, but half the class has a pretty good background. And balancing that is the challenge.}
\end{quote}
 Two educators in our data explicitly discussed this challenge. This perspective aligns with the broader discussion of QISE curriculum development, where designing QISE courses for students with varied STEM background is an issue \cite{meyer2022today}.

Amid these challenges, there were promising \textbf{opportunities} for curriculum development. Ten out of 15 educators underscored the potential of  \textit{industry-driven curriculum revisions}, emphasizing the need to align course content with emerging industry skills. For example, Fallon, a physics faculty at private research university,  stated: \begin{quote}
    \textit{So I would think carefully about, in essence, what are the modern developments that are new to QIS, that aren't currently in quantum instruction? And focus on what is it that needs to be modernized in order to make the classes more relevant for what the future lives are going to be of people who have taken them.}
\end{quote}
Another \textbf{opportunity} was \textit{sharing QISE content among educators}. Ava, who is a physics faculty at a public university, noted that her material is freely accessible to other educators: 
\begin{quote}
    \textit{So all of our materials are available freely online, and we are working to try to get the word out, in addition to developing more.}
\end{quote}
Three of our educators highlighted this opportunity to share and collaborate on QISE content development across institutions, ensuring that more instructors can get access to already developed teaching materials.

Nevertheless, curriculum development faced \textbf{threats}. Four educators expressed concern about  \textit{industry-academia misalignment on learning objectives}. Jaime explained how without clarity on industry needs, academic preparation may not align with industry desires: 
\begin{quote}
    \textit{what skills are changing and becoming more or less important as we go? I mean, I think that's often been the question with the quantum computing side of things is, what, how much, and how deeply do you need to understand, like quantum versus being able to manipulate just the gates and all of those things, versus sort of a higher level programming. How much is quantum machine learning becoming important, versus these combinations, where are those changes happening? Because I know there's always been a sense that eventually, if you're on the programming side, you probably don't deeply need to understand the quantum. I don't think we're anywhere near there yet, but that is my sense. But it'd be interesting to see from the industry side, do they still agree that most of the people they need still need to understand what's happening in terms of sort of the quantum that's going on in order to do their jobs?}
\end{quote}
 
Furthermore, two educators warned about the \textit{reliance on proprietary teaching platforms}, which can be risky. Sam shared an example: \begin{quote}
    \textit{Some other people in the program redirected their class to focus on [Company Name]’s machine, and then [Company Name] decided to no longer provide that access. [...] from my perspective that's a bad move to do, but from their perspective, they really thought they were setting up this long term.}
\end{quote}
This \textbf{threat} underscored the need for instructors to carefully choose the resources on which they base their content on to ensure sustainability. It also highlights the importance of industry to consider creating educational tools that can be maintained over the longer term, considering not only the immediate functionality, but also the ongoing impact on educators' ability to design and deliver sustainable QISE courses.

\subsubsection{Collaboration}

A notable \textbf{strength} in collaboration was working together to develop QISE education. Four educators highlighted the importance of \textit{continuous professional development}. Staying up to date on latest trends in QISE education through workshops and training was a key factor. Blake, a physics faculty at a public research university, shared an example: 
\begin{quote}
    \textit{I look out for programs at other universities or companies and see what they're doing [...] I participate in these workshops and programs and compare what I do to what they're doing to learn and improve.}
\end{quote}
This proactive professional development effort not only enhanced individual teaching practices, but also built a supportive network that is essential for the evolution of QISE education.

\subsubsection{Connections}
In the realm of external connections, one \textbf{weakness} was the \textit{limited industry connections}. Only one educator, Alex, a physics faculty at public research university, acknowledged this weakness in his course when he said: 
\begin{quote}
    \textit{Maybe industry connection is something that I need to work on, even if I don't work for the companies. I think for students, it's good to have some basic idea what they are doing, what they are looking for, so I can sort of, not entirely, but at least some partially, incorporate whatever they are looking for as a part of the class. That's something that I can work on.}
\end{quote}
This observation suggested that while educators recognize the value of industry engagement, they may not have  active connections with industry.  

Conversely, an \textbf{opportunity} in this domain was  \textit{leveraging alumni insight}. Jules reflected on interactions with former students who are now active members in the quantum industry: \begin{quote}
    \textit{I  interacted with a couple of alumni who are now working in industry with various things related to Quantum Information Sciences. I think it was really interesting to see the breakdown of jobs. The fact that most folks that are working in quantum information sciences are not really doing the quantum part themselves, there's a lot of support jobs that go into it, and that the those people would have more sort of traditional skills, right, than the folks that are really pushing the envelope in terms of developing new technologies and that sort of.}
\end{quote}
Although only one educator explicitly highlighted this factor, alumni insights offer a promising avenue for enriching course content with real-world workforce perspectives, though this may not be feasible in institutions without an established alumni network in the quantum industry. Nevertheless, connecting and understanding industry needs emerged in various ways across components of the curriculum framework, which highlights the multifaceted ways educators conceptualize input from industry in their teaching.

\subsection{Summary QISE courses}
Our analysis showed that QISE educators are navigating a range of considerations in course development. 
At the goal and vision level, institutional bureaucracy was a major hurdle, with new QISE courses often delayed by approval processes and a lack of established curricular pathways. In response, many educators were taking a pragmatic approach by embedding QISE content into existing courses. This approach not only eased implementation, but also reflected broader national approaches in QISE education \cite{Pina2025}. 

Yet, curriculum development was not without its challenges. Educators reported gaps in topic coverage and faced difficulties in designing QISE courses for students with varied STEM backgrounds. At the same time, there was momentum around aligning content with industry needs and increasing access to shared teaching materials. However, concerns remained about unclear industry expectations and reliance on proprietary tools, which can threaten long-term usability of course materials.

On the pedagogical front, some educators were embracing student-centered design that aligns with career readiness. Ongoing professional development was also another way some educators are engaging in continuous course refinement. When it comes to external engagement, direct industry collaboration was limited, though educators recognized its value. Meanwhile, alumni insights offered a promising way to connect QISE higher education training with workforce perspectives.

Notably, the Collection and Evaluation components of the curriculum framework were not explicitly mentioned by participants when discussing courses. This absence may signal that assessment practices are either assumed, underdeveloped, or not yet formalized within QISE course design.  
Overall, as QISE courses continue to develop, intentional strategies of cross-institutional collaboration, curricular standardization, and stronger industry alignment will be critical in shaping the future of QISE courses.

\subsection{QISE program development}
The educators in our study were involved in the development of QISE programs, such as degree tracks, minors, and master’s programs. Their efforts centered around setting clear program goals, structuring resources and forging partnerships that together shape a sustainable and coherent QISE education ecosystem. In our analysis, all components of the curriculum framework (Goals, Curriculum, Collaboration, Collection, Evaluation, Connections) emerged, which illustrates how each component plays a pivotal role in refining QISE programs. 

\subsubsection{Goals}

QISE program design was \textbf{strengthened} when educators approach planning with \textit{strategic program development within existing institutional structures}. This approach enabled educators to introduce new concentrations or tracks with relative ease without facing lengthy approval process. As Alex shared: 
\begin{quote}
    \textit{So we started talking about [a QISE track] earlier this year. It's been almost a year now, so it will be a concentration. That's what  I mean by track. The reason is because we already have a BS in Engineering Physics. We already have four different concentration tracks. So adding one is not going to be too hard.}
\end{quote}
This practical advantage, highlighted by eight participants, aligns with existing literature that suggests integrating QISE courses within existing STEM curricula to make QISE education more inclusive and accessible \cite{asfaw2022building}.

Another significant \textbf{strength} was the \textit{well thought-out program development} that ensures the program remains relevant. As Carter, who is the program director of a quantum program at a public research university, explained: \begin{quote}
    \textit{And so the refinements of our program have really focused on what's hard to do in a short graduate program in an academic setting.  Ideally, they would be like working with a dil[ution] fridge with a whole bunch of qubits. We want to teach them things that will be useful no matter what context they end up in. We don't want to teach them how to use this arbitrary wave generator. We want to teach them how an arbitrary wave generator works. And here are, like, multiple examples of how that gets realized, both in software and hardware [...] And so one thing we are learning for the goals of the program is that students want to learn how to do stuff.}
\end{quote} 
The learning objective about transferable skills, emphasized nine participants, demonstrated the effort to provide students with experiences that prepare them for various opportunities after they graduate.

Furthermore, the \textit{intentional interdisciplinary program development} was a notable \textbf{strength} in institutions with either dedicated quantum centers or with sufficient interdisciplinary interest in creating a QISE program. Carl, who is biology professor at a public research university and the leader of various QISE education and research projects at his institution, explained: \begin{quote}
    \textit{It would be interdisciplinary and across different schools. [...], the minor that we will do would not be residing in one department. It'll be across the whole institution, and we have similar things like that already, and it makes sense. It leverages strengths in different areas.}
\end{quote}
Six participants pointed out the value of leveraging institutional strengths from multiple disciplines when designing QISE programs. This is intertwined with the role of collaboration across  fields in developing robust QISE programs, which we will explore in more detail later in this section. 

However, some programs faced certain internal challenges. One notable \textbf{weakness} was the \textit{administrative hurdles in setting up a program}. As David, who is a physics faculty at a public research university, pointed out: \begin{quote}
    \textit{Another hurdle is then having the approval of the university. Depending on the kind of degree, there's different levels of bureaucracy and different levels of approval that are necessary. A concentration is kind of the easiest of all the different kinds of degrees we would have. But even that has a series of approvals through the deans of the colleges and the faculty senate, and someone has to do the work to shepherd those things along.}
\end{quote}
Five participants echoed similar challenges, which underscores the bureaucratic processes that complicate the creation of new programs. This challenge was essentially an expanded version of the \textit{administrative challenge of setting up a QISE course}, where the difficulties faced at the course level are scaled up to a program-wide level.

A major \textbf{opportunity} highlighted by educators was the broad \textit{exploration of quantum workforce needs}, which is framed within the high-level vision of program design. Aubrey, who is a program director of a quantum center housed in public research university, highlighted this factor: \begin{quote}
    \textit{What if a company needs  to hire an engineer? If physics is producing master's students in quantum computing, where the majority of the background of students is undergraduates in physics that go into that, will they hire them, or will they still look for the mechanical engineer or the electrical engineer?}
\end{quote}
This quote underscores the ambiguity around aligning discipline-specific academic training with industry expectations. Thirteen participants emphasized the importance of tailoring their discipline's programs to meet the specific skill set demanded by the quantum industry, which reinforces the opportunity that QISE programs have in shaping the future quantum workforce.

Another \textbf{opportunity} for educators was \textit{cross-institutional learning}, which allows them to gain insights from the experiences of other institutions in developing QISE programs. Dani, who is a physics and engineering faculty at a public research university, highlighted the value of this opportunity for less-resourced universities: 
\begin{quote}
    \textit{I think a lot of people in smaller universities or universities that don't have a lot of quantum expertise, they don't really know what the university [QISE education] landscape is, so understanding the landscape be really helpful to know.}
\end{quote}
Eight participants acknowledged this opportunity, which suggests that cross-institutional learning could strengthen the development of new QISE programs and improve their effectiveness.

One of the \textbf{threats} was the \textit{unclear utility of QISE degree programs}. Some educators question whether the anticipated demand for QISE workers will materialize as anticipated. As Fallon articulated: 
\begin{quote}
    \textit{I think the most important question,  is the demand for QIS workers really what the government is saying it is? That would be question number one. Question number two would be  what needs to happen over the next five years for QIS to become the significant enterprise that everyone is expecting it to become? And what can/might happen over the next five years to cause a collapse of the whole thing?}
\end{quote}
Seven participants expressed concerns about the future job market for QISE graduates, which is a potential threat to the long-term sustainability of standalone QISE programs.

Another \textbf{threat} came from the \textit{mismatch between academic and industry goals}. While academia focuses on providing broad skill sets  for various career paths, industry may prioritize candidates with deep, but narrow, expertise tailored to particular job roles. As Jaime explained:
\begin{quote}
    \textit{From an educator perspective, what are those broad pieces that are important? How are we training people for a breadth of possibilities? Because honestly, I think, as an educator you never want to prepare a student for one job. That's not our goal. Our goal is to educate them broadly and maybe the difference between an academic and an industry perspective on this, where they're like, we want this to go into this organization. We need to make them ready to go into a variety of places and be prepared.}
\end{quote}
Five participants noted this tension, which could complicate the alignment of educational outcomes with industry needs, thereby limiting the viability of QISE programs in the long run.  

\subsubsection{Curriculum}

A key \textbf{strength} in QISE program curricula was the \textit{broad available QISE expertise} that enables the program to cover a wide range of QISE topics. Blair, physics faculty at a public research university and the director of their quantum center, stated:
\begin{quote}
    \textit{We are fine, we have a lot of faculty in the field, so that's not an issue.}
\end{quote}
Two educators highlighted this factor, which may not reflect the reality for most institutions. In fact, institutions with fewer resources faced this \textbf{weakness} of \textit{insufficient faculty with QISE expertise}. One of our educators, Ava expressed this concern:
\begin{quote}
    \textit{We don't have the specialists in order to make it happen in a really good and authentic way.}
\end{quote}
Nine participants voiced this concern that institutions lack sufficient QISE faculty expertise to provide a robust curriculum. 

This divide in availability of QISE expertise underscores the variability in access to quantum across different institutions, which greatly influences what curriculum QISE programs can offer. To address this disparity, leveraging faculty expertise from institutions with strong QISE expertise is important. For example, dedicated QISE faculty development initiatives can help build teaching capacity at institutions with less expertise \cite{perron2021quantum}.

Moreover, \textit{lab-space constraints} were a \textbf{weakness} for some institutions, especially in creating hardware-intensive, lab-based QISE courses in their program. As Jaime highlighted: \begin{quote}
    \textit{The fact is that for us, creating hardware based programs is really hard. We have real space constraints and so, and it's hard to get equipment and so where you would need that sort of thing for a more hardware-oriented program, and that's not what we're necessarily capable of building now. }
\end{quote}
Three participants voiced this concern. While not universal, it is a  barrier for some institutions when attempting to align QISE programs with institutional resources. This points to a potential need to invest in infrastructure and resources to support the development of QISE labs.

\subsubsection{Collaboration}
Some programs had \textbf{strengths}, such as excelling in 
\textit{cross-departmental faculty collaboration}, which fosters interdisciplinary approaches to curriculum development across program course offerings. One of our interview participants, Blair discussed her approach to interdisciplinary collaboration:
\begin{quote}
    \textit{It's meant to be very interdisciplinary. We created it. So I put together a committee with representatives from six departments. So from the get go, we made it sure that people who are not traditional physics majors can take it.}
\end{quote}
Four participants discussed how QISE curricula development requires an inherently interdisciplinary approach. However, this strength was limited to institutions with the infrastructure to support interdisciplinary initiatives. 

In fact, while some institutions benefited from strong cross-departmental expertise, other institutions faced challenges in fostering interdisciplinary collaboration due to \textit{ departmental silos}. David's experience highlighted this \textbf{weakness}: 
\begin{quote}
    \textit{That's the way the concentration was originally designed. It only exists in physics right now because it was sold to the physics department by me and my colleagues in physics. There's one chemist  and she hasn't really sold it to her department and the EE department. Although there are courses taught in those electives that our physics students can use, they don't have the existing concentration in the same way. So part of it the hurdle, is getting the departments to agree to this, and having the faculty who do the work to get it done}
\end{quote}
Four participants mentioned this issue, which highlights how program development is often driven by individual faculty within their respective departments. While some institutions highlighted faculty collaboration as a strength, others faced the difficulty of aligning different disciplines under a shared QISE vision.

\subsubsection{Collection}
Some  QISE programs benefited from some level of \textit{longitudinal data gathering}.  These efforts provided valuable insights into student progress and program impact. For example, Carl, discussed the use of pre- and post-surveys to measure student progress: 
\begin{quote}
    \textit{We actually designed a short survey that we do pre and post and testing of the participants. So basically, there's some basic stuff that in a variety of domains that we ask the students to gauge where they are at the beginning, and then at the end, we redo that survey and we determine pre and post differences. And there were significant gains after just one semester for that general introduction for all.  So we felt that for the goal of increasing awareness, we were able to accomplish that with the design of the course as it was, not requiring them to have taken Python, not requiring them to have taken linear algebra.}
\end{quote}
Three educators mentioned data collection processes in their QISE efforts. This structured approach to data collection allowed program directors and faculty to measure learning outcomes and program effectiveness.

However, having only \textit{anecdotal data} remained a common approach, as Blair reflected, 
\begin{quote}
    \textit{I would say that content is not really changing that much. What we change is the pace, what we emphasize. So, one thing that, so I've taught it twice, and other instructors have taught it one time each, and the last feedback I heard from the person who's teaching it now is that one thing they seem to have trouble with is measurement in different bases, which is an advanced thing. It's not surprising. The feedback so far is more like, what should we emphasize more?}
\end{quote}
Two participants mentioned using anecdotal feedback to inform changes. This suggests that while some programs rely on structured data to inform improvements, others use less formal, anecdotal feedback, which limits the ability to track long-term outcomes.

\subsubsection{Evaluation}

\textit{Summative assessment} was a \textbf{strength} for evaluating QISE program effectiveness. Aubrey outlined her approach: \begin{quote}
    \textit{In my main role, I look at statistics outcomes, I look at what the students are telling us, where they want to end up, how that changes throughout the program, and I just look at the outcomes, and also keep asking the students for feedback, what they like about the program, what they don't like about the program, what they think would make it stronger, while at the same time, interact with our faculty and bounce ideas on how we can change the curriculum, or what can we do to strengthen the program}
\end{quote}
Two participants mentioned conducting program assessments, highlighting a potential gap in systematic program evaluation. 

An \textbf{opportunity} arose in \textit{evaluation of outcomes data from other programs} to understand broader trends and best practices.  As Anna, a physics faculty at a private university, highlighted:
\begin{quote}
    \textit{I think there's going to be great interest in the longitudinal study of what have students going out of different programs done, you know, what? What jobs have they gone to? This is really early on. But are these programs keeping track, kind of like the APS studies, where physics bachelor's go? One of the things that I want to stress that's so helpful when we're trying to argue with a dean to support this minor or a track in the major, is the fact that you're publishing in Phys[ical] Rev[iew] and giving us these statistics is enormously helpful.}
\end{quote}
Two participants underscored this opportunity, which emphasizes the potential benefits of gathering comparative data across institutions to inform future program development.  Data gathering could help inform program development, particularly in helping individual programs position themselves within the broader landscape of QISE education.

\subsubsection{Connections}
Some institutions had successfully  \textit{established partnerships with quantum companies}, which is a \textbf{strength} because it offers students the opportunity to gain practical, real-world experience in the quantum industry. As Carl explained: \begin{quote}
    \textit{We do work with the [Company Name] quantum center, and which was the entity that got us engaged in this, in building, really diversifying the QISE workforce, in particular, trying to get more HBCUs students in particular, African American students involved in QISE related activities. And that goes from basic training in principles of QISE versus engaging them in co-curricular activities, such for everything from hackathons to internships and engaging them in academic year research activities with faculty.}
\end{quote} 
Five educators mentioned having strong partnerships with industry, which provides students with the opportunity to get hands-on experience in the quantum industry. While not universally available, these partnerships point to a viable path for institutions seeking to deepen industry engagement. 

\subsection{Summary QISE programs}
The development of QISE programs involved designing degree tracks and curricula that align with industry needs while leveraging institutional strengths. A major strength lied in integrating QISE within existing institutional structures, such as the broad available QISE expertise and intentional interdisciplinary program development, which streamlined the process of launching new programs. However, administrative hurdles and space constraints, particularly for lab courses,  can slow down program development. There was also uncertainty regarding the demand for QISE graduates with master's degrees, which poses a potential threat to the long-term viability of standalone QISE programs. 

The availability of broad QISE expertise at institutions that are well-resources in QISE enabled some programs to cover a wide range of topics. However, insufficient faculty expertise was a challenge for other institutions with less QISE-specific resources, leading to variability in curriculum offerings. This disparity highlighted the need for coordinated curriculum and collaboration to support less-resourced programs \cite{barnes2025outcomes}.

Successful cross-departmental collaboration strengthened QISE curricula, though this was hindered by departmental silos in some cases. Partnerships with quantum companies also provided students with practical experience, although such connections were not universal.

Program-level data collection practices varied. Some institutions used surveys to measure program outcomes, while others relied on anecdotal feedback, which limited the ability to track long-term effectiveness. Summative assessments and comparative data across institutions provided valuable insights for refining QISE programs, but these practices were not widespread, indicating a gap in systematic evaluation.

Overall, QISE program development was shaped by a dynamic interplay of institutional strengths, resource availability, and an evolving quantum industry landscape. Addressing these challenges through strategic coordination, targeted investment, and systematic evaluation is necessary to build sustainable QISE programs.

\section{Discussion}\label{discussion}
Our study set out to answer these two research questions: \begin{itemize}
    \item How are educators designing, implementing, and sustaining QISE programs and courses in an effort to better support students’ education?
    \item What insights do educators' experiences offer regarding the future development of QISE education?
\end{itemize} Through our analysis, we identified a range of internal and external factors that educators navigate, depending on the focus of their QISE efforts (courses or programs).  Table \ref{tab:results} provides an overview of the most prominent considerations in the minds of educators who are deeply engaged in QISE education.  
Across all factors, five overarching themes emerged, each tied to a component of the curricular framework, and allowed us to draw insights about the future development of QISE education. 

Although combining a curriculum framework with a SWOT analysis was particularly valuable in identifying key themes, one limitation of this approach was that some framework components lacked associated SWOT factors as seen in Table \ref{tab:results}.  This was potentially due to the constraints of doing SWOT at a community level across multiple institutions. However, this presents an opportunity for educators to thoughtfully consider these areas as they develop their courses and programs, potentially identifying overlooked aspects that could strengthen their educational efforts. For example, as we will show in the themes in this section, we combined  the collection and evaluation components, as they form a single cohesive theme that showcases a potentially overlooked aspect in QISE course and program design. 

\subsection{Goals: Navigating Institutional Structures for QISE Course and Program Development}

QISE educators work within existing institutional structures to establish and sustain QISE courses and programs.  As a strategic approach, many educators choose to integrate QISE topics into existing courses or introduce QISE courses as additions to existing degree programs. This approach streamlines bureaucratic approval processes and minimizes the need for additional financial or physical resources. This practice is scalable and accessible, regardless of institution type or available resources, and thus can be used by faculty as they get started in QISE. 

Depending on institutional resources, some educators capitalize on broad cross-disciplinary interest to create an interdisciplinary program that aligns with workforce demands. However, the evolving and uncertain nature of quantum industry needs make it challenging to define clear and long-term objectives of QISE courses and programs. These variations highlight how institutional context directly shapes the design and scope of QISE offerings.

This theme aligns with established instructional design models, which emphasize the importance of clearly defining the goals of a course or a program during development \cite{chasteen2011thoughtful}. However, what sets QISE education apart is that it is shaped by uncertain workforce needs and lacks a singular disciplinary home. In this context, setting high-level goals for courses and programs is a moving and dynamic challenge. Hence, the broader institutional and disciplinary context play a critical role in shaping the direction and goals of QISE education.

\subsection{Curriculum: Aligning QISE Curriculum Development with Industry and Institutional Needs}

The lack of a standardized curriculum in QISE presents unique challenges in course and program development. Educators navigate several hurdles, including designing QISE courses for varied STEM backgrounds, lacking sufficient specialized expertise to cover the depth and breath of QISE, and identifying which industry-relevant skills or content to prioritize. In response, educators are encouraging the sharing of content across institutions and incorporating industry input to refine materials and establish workforce needs. 

Some of the curriculum development challenges highlighted in our analysis align with prior work. In particular, the challenges of developing courses for varying levels of students’ preparation and the lack of faculty with QISE expertise \cite{meyer2022today} remain persistent barriers for faculty. The nuances in this theme provide further evidence to support what leaders in the quantum education space are advocating for: a national quantum center to coordinate and expand QISE curriculum across institution types, particularly institutions with varying levels of existing QISE activity \cite{barnes2025outcomes}. Among other roles to support quantum education, such a center could serve as a hub for resource sharing and the alignment of QISE curriculum development with industry needs. Inspiration and best practices could be drawn from other physics education resource efforts that have been built over the years, including but not limited to PhysPort \cite{mckagan2020physport} and the Living Physics Portal \cite{chessey2023living}.

\subsection{Collaboration : Fostering Faculty Partnerships within and across Institutions to Strengthen QISE Education}

Collaboration is central to the development of QISE education, especially for program development. 
Although some institutions with QISE centers have established effective cross-faculty collaboration, others face challenges due to departmental silos that hinder interdisciplinary courses and program development. Additionally, because QISE expertise remains concentrated within institutions that are well-resourced in QISE, cross-institutional partnerships are essential for supporting faculty at all types of institutions. Some educators engage in professional development to be part of a growing and supportive QISE education community. These collaborative efforts are key to designing, implementing, and sustaining QISE courses and programs that reflect the field’s growth and interdisciplinary nature.

Collaboration is not only foundational to the success of one QISE course or program, but also crucial for advancing the QISE education landscape. While QISE is inherently interdisciplinary, our study highlights that there is need for deeper insight into how to foster more effective cross-disciplinary faculty collaboration. Expanding collaboration across departments and institutions will ensure a more cohesive QISE education system. This is another example of how a national quantum center, which supports coordinated activities across QISE education, would support the collaboration needs of educators across various institutions \cite{barnes2025outcomes}. Enhancing collaboration at both the institution and national levels is essential for creating a robust QISE education ecosystem that benefits all stakeholders. 

\subsection{Data Collection and Evaluation: Need for Systematic Approaches to Guide QISE Courses and Programs Improvement}

Our results showcase the need for educators to be more intentional about the data they collect to assess the effectiveness of their QISE courses and programs. Even though QISE courses and programs are still relatively new,  gathering data and evaluating success of courses and programs would strengthen their development. While some programs have established longitudinal data collection efforts, others rely solely on anecdotal evidence.  In the early stages of QISE education development, it would be an opportune time to consider the data needed to collect to sustain QISE courses and programs. 

In contrast to other areas within physics that benefit from established evidence-based assessments \cite{hestenes1992force, Madsen2017RBAI1, Madsen2019RBAI2}, QISE courses currently lack systematic tools to measure student learning  and alignment with workforce needs. Although there are some established quantum mechanics assessments \cite{cataloglu2002testing, wuttiprom2009development, mckagan2010design, sadaghiani2015quantum, marshman2019validation}, QISE-specific assessments are still under development \cite{meyer2024introductory, Durkin2025QISCIT}. For program-level evaluation, existing frameworks (e.g., Ref.~\cite{ABETAccreditation, HLC}), and existing guides such as the Effective Practices for Physics Programs  \cite{ep3_assess_learning} could be useful starting points that could be adapted to meet the specific needs of quantum information science programs. One potential approach to consolidate efforts  is also through a national quantum center \cite{barnes2025outcomes} that could coordinate broad data collection and analysis on QISE education. Ultimately, expanding data-driven efforts across institutions will help ensure that QISE education is responsive to students and workforce needs.

\subsection{Connections: Strengthening Industry and Academic Connections to Prepare Students for Workforce}
Both courses and programs face challenges related to understanding the quantum job market and the alignment between academic preparation and workforce expectations. While some programs have strong industry partnerships, others have limited connections to quantum companies. Strengthening academic-industry collaborations and alumni engagement will help students navigate career opportunities, while also ensuring QISE curricula align with workforce needs. One possible way to strengthen these collaborations across all types of institutions would be through regional partnerships, which can provide greater access to industry connections and resources, which benefits all stakeholders.  

Although some work has been done in bridging the gap between higher education and quantum industry \cite{fox2020preparing, aiello2021achieving, kaur2022defining, hughes2022assessing, hasanovic2022quantum}, our educators highlight that there continues to be a need to better understand the quantum workforce to help guide improvements to QISE education and make opportunities more accessible and helpful for students who want to enter the workforce. The latter is the subject of our research team’s ongoing work to improve QISE education and prepare interested students for the quantum workforce \cite{RIT_Industry_Perspectives}. By collecting and sharing industry data, our work, as well as that of others, can support educators regardless of their institution's level of resources, which will help inform curriculum development and program strategies.  Ultimately, fostering stronger industry ties and workforce insights will create a more cohesive and impactful QISE education experience for students. 

\section{Conclusion}
To conclude, we combined a SWOT analysis with a curriculum framework to provide an overarching perspective of the considerations educators are dealing with as they design and implement their QISE courses and programs. The curriculum framework used in this work provides a structured approach for educators to design and develop QISE courses and programs. Although there is no universal model for implementation, this framework encourages maintaining a broad perspective, while tailoring efforts to specific institutional contexts. It outlines key components to consider and when paired with a SWOT analysis, helps to elucidate the opportunities and challenges within each area. 

Although we applied this approach at the community level, individual educators can adapt it to their own context. Program directors and faculty can use it to reflect on their local context, identify relevant factors, and make informed decisions in QISE course and program development. This approach not only supports more intentional curriculum design, but also promotes alignment between QISE education and opportunities in the quantum workforce. 
Our results illustrate the dynamic landscape of factors shaping QISE education, influenced by instructional structures, curriculum development, collaboration among educators, program data collection and evaluation, and industry connections. The nuances and interconnections in our findings illustrate the value of applying a structured approach to QISE course and program development in order to support educators in creating a more cohesive QISE education ecosystem.
 
\begin{acknowledgments}
The authors thank the members of the Lewandowski physics education research group for their feedback on this project. This work is based on work supported by the National Science Foundation under Grant Nos. PHY-2333073 and PHY-2333074. 
\end{acknowledgments}


\onecolumngrid
\pagebreak

\appendix

\section{Operational definitions of SWOT in curriculum framework} \label{appendix:swot_definitions}

\begin{table*}[h!]
\centering
\caption{Operational SWOT definitions within each component of the framework}
\begin{tabular}{p{17cm}}
\toprule
\multicolumn{1}{c}{} \\[-0.5em]

\textbf{Goals} 
\begin{itemize}
    \item[] Strengths: Internal structural considerations that ensure a clear high-level vision and outcomes for QISE courses and programs
    \item[] Weaknesses: Internal structural considerations that could limit implementing a QISE course or program
    \item[] Opportunities: External trends or developments that support creation and refinement of QISE course or program
    \item[] Threats: External factors that challenge the sustainability of a clear vision and structure for a QISE course or program
\end{itemize}
\\ \hline
\textbf{Curriculum}
\begin{itemize}
    \item[] Strengths: Internal factors that facilitate QISE curriculum integration to institutions' unique resources and context
    \item[] Weaknesses: Internal structural considerations that could limit implementing QISE curriculum
    \item[] Opportunities: External factors that enable better alignment of QISE curriculum with the institution's unique context 
    \item[] Threats: External factors that limit the ability to effectively adapt QISE curriculum to the institution's context
\end{itemize}
\\ \hline
\textbf{Collaboration}
\begin{itemize}
    \item[] Strengths: Strong collaborations across disciplines and departments that can provide interdisciplinary curriculum for students
    \item[] Weaknesses: Inefficiencies in collaboration among faculty within and across departments
    \item[] Opportunities: External factors that encourage stronger collaboration and alignment across stakeholders
    \item[] Threats: External factors that hinder collaboration and alignment among stakeholders
\end{itemize}
\\ \hline
\textbf{Collection}
\begin{itemize}
    \item[] Strengths: Effective processes and tools for systematically gathering evidence of student learning
    \item[] Weaknesses: Gaps in processes or tools used for systematic data collection on students' academic performance in QISE courses or programs
    \item[] Opportunities: External factors that support the development of effective data collection about students in QISE courses or programs 
    \item[] Threats: External barriers to systematically gathering data on student progress in QISE courses and programs
\end{itemize}
\\ \hline
\textbf{Evaluation}
\begin{itemize}
    \item[] Strengths: A well-developed approach to aligning academic assessments with industry-valued skills to ensure graduates are well prepared
    \item[] Weaknesses: Internal barriers to having a comprehensive evaluation of students in courses or programs
    \item[] Opportunities: External factors that support comprehensive assessment of data collected
    \item[] Threats: External factors challenging the ability to create a comprehensive evaluation of courses or programs
\end{itemize}
\\ \hline
\textbf{Connections}
\begin{itemize}
    \item[] Strengths: Existing connections with industry that enhance the program's relevance and impact 
    \item[] Weaknesses: Internal shortcomings in establishing or maintaining effective connections between academia and industry 
    \item[] Opportunities: Avenues for new or expanded collaboration between academia and industry to refine course or program offerings and align with job-market needs
    \item[] Threats: External issues that would disrupt collaboration and feedback loops between academic and industry stakeholders
\end{itemize}
\\ \hline
\hline
\end{tabular}
\label{tab:swot_definitions}
\end{table*}

\section{Interview participants}\label{appendix:interview_participants}
\begin{table*}[h!]
\centering
\renewcommand{\arraystretch}{1.6} 
\setlength{\tabcolsep}{8pt} 
\caption{Interview participants, institution type, primary department, type of QISE effort (course or program), and QISE specific institutional resources. MSI stands for Minority Serving Institutions. HSI stands for Hispanic Serving Institutions. HBCU stands for Historically Black Colleges and Universities. QISE courses refer to a standalone QISE course or added QISE topics to an existing quantum course. QISE programs refer to either concentration or track, minor or master's program. Design means that the planning for the course or program started, but has not been implemented yet. Implementation means that the course or program was developed and is ongoing.}

\begin{tabular}{
    >{\centering\arraybackslash}p{1.5cm}
    >{\centering\arraybackslash}p{4.5cm}
    >{\centering\arraybackslash}p{2cm}
    >{\centering\arraybackslash}p{5.2cm}
    >{\centering\arraybackslash}p{1.8cm}
}
\toprule
\textbf{Pseudonym} & \textbf{Institution type} & \textbf{Home department} & \textbf{QISE effort and development stage} & \textbf{QISE institutional resources} \\ \hline
\hline 
Jules & Doctoral/Professional Universities, Non-MSI, Private & Physics & Non-QISE course with QISE topics: implementation \newline Minor in QISE: design & -- \\ \hline
Casey & Master's Colleges and Universities: Larger Programs, MSI, HSI, Public & Physics & QISE course: implementation \newline Non-QISE course with QISE topic: implementation & -- \\ \hline
Ava & High Research Activity, MSI, HSI, Public & Physics & QISE course: implementation & -- \\ \hline
Anna & Doctoral/Professional Universities, Non-MSI, Private & Physics & Non-QISE course with QISE topics: implementation \newline Minor in QISE: design & -- \\ \hline
Blake & Very High Research Activity, HSI, Public & Physics & QISE course: implementation & -- \\ \hline
Alex & Very High Research Activity, Non-MSI, Public & Physics & QISE course: implementation & -- \\ \hline
Fallon & Very High Research Activity, Non-MSI, Private & Physics & QISE course: implementation & -- \\ \hline
Carl & High Research Activity, MSI, HBCU, Public & Biology & QISE course: implementation \newline Minor in QISE: design & -- \\ \hline
David &  Very High Research Activity, HSI, Public & Physics & QISE course: implementation \newline PhD concentration in QISE: design & Quantum institute \\ \hline
Sam & Very High Research Activity, Non-MSI, Private & Engineering & QISE course: implementation; Master's concentration in QISE program: implementation & QISE center \\ \hline
Dani & Very High Research Activity, Non-MSI, Public & Physics and Engineering & QISE course: implementation Quantum engineering program: implementation & Quantum hub \\ \hline
Jaime & Very High Research Activity, Non-MSI, Public & Physics & QISE course: implementation \newline Master's concentration in QISE: implementation \newline Master's in QISE: design & QISE center \\ \hline
Aubrey & Very High Research Activity, Non-MSI, Public & Physics & QISE master's program: implementation & Quantum institute \\ \hline
Carter & Very High Research Activity, Non-MSI, Public & Physics & QISE course: implementation \newline QISE master's program: implementation & QISE center \\ \hline
Blair & Very High Research Activity, Non-MSI, Public & Physics & QISE course: implementation \newline QISE master's program: implementation & QISE center \\
\hline \hline
\end{tabular}
\label{tab:interview_participants}
\end{table*}

\twocolumngrid
\pagebreak

\bibliography{apssamp}

\begin{thebibliography}{60}%
\makeatletter
\providecommand \@ifxundefined [1]{%
 \@ifx{#1\undefined}
}%
\providecommand \@ifnum [1]{%
 \ifnum #1\expandafter \@firstoftwo
 \else \expandafter \@secondoftwo
 \fi
}%
\providecommand \@ifx [1]{%
 \ifx #1\expandafter \@firstoftwo
 \else \expandafter \@secondoftwo
 \fi
}%
\providecommand \natexlab [1]{#1}%
\providecommand \enquote  [1]{``#1''}%
\providecommand \bibnamefont  [1]{#1}%
\providecommand \bibfnamefont [1]{#1}%
\providecommand \citenamefont [1]{#1}%
\providecommand \href@noop [0]{\@secondoftwo}%
\providecommand \href [0]{\begingroup \@sanitize@url \@href}%
\providecommand \@href[1]{\@@startlink{#1}\@@href}%
\providecommand \@@href[1]{\endgroup#1\@@endlink}%
\providecommand \@sanitize@url [0]{\catcode `\\12\catcode `\$12\catcode `\&12\catcode `\#12\catcode `\^12\catcode `\_12\catcode `\%12\relax}%
\providecommand \@@startlink[1]{}%
\providecommand \@@endlink[0]{}%
\providecommand \url  [0]{\begingroup\@sanitize@url \@url }%
\providecommand \@url [1]{\endgroup\@href {#1}{\urlprefix }}%
\providecommand \urlprefix  [0]{URL }%
\providecommand \Eprint [0]{\href }%
\providecommand \doibase [0]{https://doi.org/}%
\providecommand \selectlanguage [0]{\@gobble}%
\providecommand \bibinfo  [0]{\@secondoftwo}%
\providecommand \bibfield  [0]{\@secondoftwo}%
\providecommand \translation [1]{[#1]}%
\providecommand \BibitemOpen [0]{}%
\providecommand \bibitemStop [0]{}%
\providecommand \bibitemNoStop [0]{.\EOS\space}%
\providecommand \EOS [0]{\spacefactor3000\relax}%
\providecommand \BibitemShut  [1]{\csname bibitem#1\endcsname}%
\let\auto@bib@innerbib\@empty
\bibitem [{\citenamefont {{National Quantum Initiative}}(2018)}]{NQI}%
  \BibitemOpen
  \bibfield  {author} {\bibinfo {author} {\bibnamefont {{National Quantum Initiative}}},\ }\href@noop {} {\bibinfo {title} {National quantum initiative}},\ \bibinfo {howpublished} {\url{http://quantum.gov/}} (\bibinfo {year} {2018})\BibitemShut {NoStop}%
\bibitem [{\citenamefont {Riedel}\ \emph {et~al.}(2019)\citenamefont {Riedel}, \citenamefont {Kovacs}, \citenamefont {Zoller}, \citenamefont {Mlynek},\ and\ \citenamefont {Calarco}}]{riedel2019europe}%
  \BibitemOpen
  \bibfield  {author} {\bibinfo {author} {\bibfnamefont {M.}~\bibnamefont {Riedel}}, \bibinfo {author} {\bibfnamefont {M.}~\bibnamefont {Kovacs}}, \bibinfo {author} {\bibfnamefont {P.}~\bibnamefont {Zoller}}, \bibinfo {author} {\bibfnamefont {J.}~\bibnamefont {Mlynek}},\ and\ \bibinfo {author} {\bibfnamefont {T.}~\bibnamefont {Calarco}},\ }\bibfield  {title} {\bibinfo {title} {Europe’s quantum flagship initiative},\ }\href@noop {} {\bibfield  {journal} {\bibinfo  {journal} {Quantum Science and Technology}\ }\textbf {\bibinfo {volume} {4}},\ \bibinfo {pages} {020501} (\bibinfo {year} {2019})}\BibitemShut {NoStop}%
\bibitem [{\citenamefont {Knight}\ and\ \citenamefont {Walmsley}(2019)}]{knight2019uk}%
  \BibitemOpen
  \bibfield  {author} {\bibinfo {author} {\bibfnamefont {P.}~\bibnamefont {Knight}}\ and\ \bibinfo {author} {\bibfnamefont {I.}~\bibnamefont {Walmsley}},\ }\bibfield  {title} {\bibinfo {title} {Uk national quantum technology programme},\ }\href@noop {} {\bibfield  {journal} {\bibinfo  {journal} {Quantum Science and Technology}\ }\textbf {\bibinfo {volume} {4}},\ \bibinfo {pages} {040502} (\bibinfo {year} {2019})}\BibitemShut {NoStop}%
\bibitem [{\citenamefont {Subcommittee~on Quantum Information~Science}\ and\ \citenamefont {Council}(2022)}]{NSP}%
  \BibitemOpen
  \bibfield  {author} {\bibinfo {author} {\bibfnamefont {N.~S.}\ \bibnamefont {Subcommittee~on Quantum Information~Science}}\ and\ \bibinfo {author} {\bibfnamefont {T.}~\bibnamefont {Council}},\ }\href {https://www.quantum.gov/wp-content/uploads/2022/02/QIST-Natl-Workforce-Plan.pdf} {\bibinfo {title} {National strategic plan for quantum information science and technology workforce}} (\bibinfo {year} {2022})\BibitemShut {NoStop}%
\bibitem [{\citenamefont {Congress}(2022)}]{CHIPS}%
  \BibitemOpen
  \bibfield  {author} {\bibinfo {author} {\bibfnamefont {U.}~\bibnamefont {Congress}},\ }\href {https://www.congress.gov/bill/117th-congress/house-bill/4346} {\bibinfo {title} {{CHIPS} and science act of 2022}} (\bibinfo {year} {2022}),\ \bibinfo {note} {public Law No: 117-167}\BibitemShut {NoStop}%
\bibitem [{\citenamefont {Taylor}(2023)}]{taylor2023us}%
  \BibitemOpen
  \bibfield  {author} {\bibinfo {author} {\bibfnamefont {M.}~\bibnamefont {Taylor}},\ }\href@noop {} {\bibinfo {title} {The {US} {CHIPS} and science act of 2022}} (\bibinfo {year} {2023})\BibitemShut {NoStop}%
\bibitem [{\citenamefont {Perron}\ \emph {et~al.}(2021)\citenamefont {Perron}, \citenamefont {DeLeone}, \citenamefont {Sharif}, \citenamefont {Carter}, \citenamefont {Grossman}, \citenamefont {Passante},\ and\ \citenamefont {Sack}}]{perron2021quantum}%
  \BibitemOpen
  \bibfield  {author} {\bibinfo {author} {\bibfnamefont {J.~K.}\ \bibnamefont {Perron}}, \bibinfo {author} {\bibfnamefont {C.}~\bibnamefont {DeLeone}}, \bibinfo {author} {\bibfnamefont {S.}~\bibnamefont {Sharif}}, \bibinfo {author} {\bibfnamefont {T.}~\bibnamefont {Carter}}, \bibinfo {author} {\bibfnamefont {J.~M.}\ \bibnamefont {Grossman}}, \bibinfo {author} {\bibfnamefont {G.}~\bibnamefont {Passante}},\ and\ \bibinfo {author} {\bibfnamefont {J.}~\bibnamefont {Sack}},\ }\bibfield  {title} {\bibinfo {title} {Quantum undergraduate education and scientific training},\ }\href@noop {} {\bibfield  {journal} {\bibinfo  {journal} {arXiv preprint arXiv:2109.13850}\ } (\bibinfo {year} {2021})}\BibitemShut {NoStop}%
\bibitem [{\citenamefont {Economou}\ and\ \citenamefont {Barnes}(2022)}]{economou2022hello}%
  \BibitemOpen
  \bibfield  {author} {\bibinfo {author} {\bibfnamefont {S.~E.}\ \bibnamefont {Economou}}\ and\ \bibinfo {author} {\bibfnamefont {E.}~\bibnamefont {Barnes}},\ }\bibfield  {title} {\bibinfo {title} {Hello quantum world! a rigorous but accessible first-year university course in quantum information science},\ }\href@noop {} {\bibfield  {journal} {\bibinfo  {journal} {arXiv preprint arXiv:2210.02868}\ } (\bibinfo {year} {2022})}\BibitemShut {NoStop}%
\bibitem [{\citenamefont {Asfaw}\ \emph {et~al.}(2022)\citenamefont {Asfaw}, \citenamefont {Blais}, \citenamefont {Brown}, \citenamefont {Candelaria}, \citenamefont {Cantwell}, \citenamefont {Carr}, \citenamefont {Combes}, \citenamefont {Debroy}, \citenamefont {Donohue}, \citenamefont {Economou} \emph {et~al.}}]{asfaw2022building}%
  \BibitemOpen
  \bibfield  {author} {\bibinfo {author} {\bibfnamefont {A.}~\bibnamefont {Asfaw}}, \bibinfo {author} {\bibfnamefont {A.}~\bibnamefont {Blais}}, \bibinfo {author} {\bibfnamefont {K.~R.}\ \bibnamefont {Brown}}, \bibinfo {author} {\bibfnamefont {J.}~\bibnamefont {Candelaria}}, \bibinfo {author} {\bibfnamefont {C.}~\bibnamefont {Cantwell}}, \bibinfo {author} {\bibfnamefont {L.~D.}\ \bibnamefont {Carr}}, \bibinfo {author} {\bibfnamefont {J.}~\bibnamefont {Combes}}, \bibinfo {author} {\bibfnamefont {D.~M.}\ \bibnamefont {Debroy}}, \bibinfo {author} {\bibfnamefont {J.~M.}\ \bibnamefont {Donohue}}, \bibinfo {author} {\bibfnamefont {S.~E.}\ \bibnamefont {Economou}}, \emph {et~al.},\ }\bibfield  {title} {\bibinfo {title} {Building a quantum engineering undergraduate program},\ }\href@noop {} {\bibfield  {journal} {\bibinfo  {journal} {IEEE Transactions on Education}\ }\textbf {\bibinfo {volume} {65}},\ \bibinfo {pages} {220} (\bibinfo {year} {2022})}\BibitemShut {NoStop}%
\bibitem [{\citenamefont {Blanchette}\ \emph {et~al.}(2024)\citenamefont {Blanchette}, \citenamefont {Normandin}, \citenamefont {Pioro-Ladri{\`e}re}, \citenamefont {St-Hilaire}, \citenamefont {Soldera}, \citenamefont {Thibault},\ and\ \citenamefont {Touchette}}]{blanchette2024design}%
  \BibitemOpen
  \bibfield  {author} {\bibinfo {author} {\bibfnamefont {S.}~\bibnamefont {Blanchette}}, \bibinfo {author} {\bibfnamefont {D.}~\bibnamefont {Normandin}}, \bibinfo {author} {\bibfnamefont {M.}~\bibnamefont {Pioro-Ladri{\`e}re}}, \bibinfo {author} {\bibfnamefont {L.}~\bibnamefont {St-Hilaire}}, \bibinfo {author} {\bibfnamefont {A.}~\bibnamefont {Soldera}}, \bibinfo {author} {\bibfnamefont {K.}~\bibnamefont {Thibault}},\ and\ \bibinfo {author} {\bibfnamefont {D.}~\bibnamefont {Touchette}},\ }\bibfield  {title} {\bibinfo {title} {The design and implementation of a quantum information science undergraduate program},\ }\href@noop {} {\bibfield  {journal} {\bibinfo  {journal} {arXiv preprint arXiv:2412.01874}\ } (\bibinfo {year} {2024})}\BibitemShut {NoStop}%
\bibitem [{\citenamefont {Goorney}\ \emph {et~al.}(2024{\natexlab{a}})\citenamefont {Goorney}, \citenamefont {Sarantinou},\ and\ \citenamefont {Sherson}}]{goorney2024quantum}%
  \BibitemOpen
  \bibfield  {author} {\bibinfo {author} {\bibfnamefont {S.}~\bibnamefont {Goorney}}, \bibinfo {author} {\bibfnamefont {M.}~\bibnamefont {Sarantinou}},\ and\ \bibinfo {author} {\bibfnamefont {J.}~\bibnamefont {Sherson}},\ }\bibfield  {title} {\bibinfo {title} {The quantum technology open master: widening access to the quantum industry},\ }\href@noop {} {\bibfield  {journal} {\bibinfo  {journal} {EPJ Quantum Technology}\ }\textbf {\bibinfo {volume} {11}},\ \bibinfo {pages} {7} (\bibinfo {year} {2024}{\natexlab{a}})}\BibitemShut {NoStop}%
\bibitem [{\citenamefont {Qerimi}\ \emph {et~al.}(2025)\citenamefont {Qerimi}, \citenamefont {Malone}, \citenamefont {Rexigel}, \citenamefont {Mehlhase}, \citenamefont {Kuhn},\ and\ \citenamefont {Küchemann}}]{qerimi2025exploring}%
  \BibitemOpen
  \bibfield  {author} {\bibinfo {author} {\bibfnamefont {L.}~\bibnamefont {Qerimi}}, \bibinfo {author} {\bibfnamefont {S.}~\bibnamefont {Malone}}, \bibinfo {author} {\bibfnamefont {E.}~\bibnamefont {Rexigel}}, \bibinfo {author} {\bibfnamefont {S.}~\bibnamefont {Mehlhase}}, \bibinfo {author} {\bibfnamefont {J.}~\bibnamefont {Kuhn}},\ and\ \bibinfo {author} {\bibfnamefont {S.}~\bibnamefont {Küchemann}},\ }\bibfield  {title} {\bibinfo {title} {Exploring the mechanisms of qubit representations and introducing a new category system for visual representations: results from expert ratings},\ }\bibfield  {journal} {\bibinfo  {journal} {EPJ Quantum Technology}\ }\textbf {\bibinfo {volume} {12}},\ \href {https://doi.org/10.1140/epjqt/s40507-025-00091-1} {10.1140/epjqt/s40507-025-00091-1} (\bibinfo {year} {2025})\BibitemShut {NoStop}%
\bibitem [{\citenamefont {DeVore}\ and\ \citenamefont {Singh}(2016)}]{devore2016development}%
  \BibitemOpen
  \bibfield  {author} {\bibinfo {author} {\bibfnamefont {S.}~\bibnamefont {DeVore}}\ and\ \bibinfo {author} {\bibfnamefont {C.}~\bibnamefont {Singh}},\ }\bibfield  {title} {\bibinfo {title} {Development of an interactive tutorial on quantum key distribution},\ }\href@noop {} {\bibfield  {journal} {\bibinfo  {journal} {arXiv preprint arXiv:1601.00730}\ } (\bibinfo {year} {2016})}\BibitemShut {NoStop}%
\bibitem [{\citenamefont {Economou}\ \emph {et~al.}(2020)\citenamefont {Economou}, \citenamefont {Rudolph},\ and\ \citenamefont {Barnes}}]{economou2020teaching}%
  \BibitemOpen
  \bibfield  {author} {\bibinfo {author} {\bibfnamefont {S.~E.}\ \bibnamefont {Economou}}, \bibinfo {author} {\bibfnamefont {T.}~\bibnamefont {Rudolph}},\ and\ \bibinfo {author} {\bibfnamefont {E.}~\bibnamefont {Barnes}},\ }\bibfield  {title} {\bibinfo {title} {Teaching quantum information science to high-school and early undergraduate students},\ }\href@noop {} {\bibfield  {journal} {\bibinfo  {journal} {arXiv preprint arXiv:2005.07874}\ } (\bibinfo {year} {2020})}\BibitemShut {NoStop}%
\bibitem [{\citenamefont {Salehi}\ \emph {et~al.}(2021)\citenamefont {Salehi}, \citenamefont {Seskir},\ and\ \citenamefont {Tepe}}]{salehi2021computer}%
  \BibitemOpen
  \bibfield  {author} {\bibinfo {author} {\bibfnamefont {{\"O}.}~\bibnamefont {Salehi}}, \bibinfo {author} {\bibfnamefont {Z.}~\bibnamefont {Seskir}},\ and\ \bibinfo {author} {\bibfnamefont {I.}~\bibnamefont {Tepe}},\ }\bibfield  {title} {\bibinfo {title} {A computer science-oriented approach to introduce quantum computing to a new audience},\ }\href@noop {} {\bibfield  {journal} {\bibinfo  {journal} {IEEE Transactions on Education}\ }\textbf {\bibinfo {volume} {65}},\ \bibinfo {pages} {1} (\bibinfo {year} {2021})}\BibitemShut {NoStop}%
\bibitem [{\citenamefont {Cervantes}\ \emph {et~al.}(2021)\citenamefont {Cervantes}, \citenamefont {Passante}, \citenamefont {Wilcox},\ and\ \citenamefont {Pollock}}]{cervantes2021overview}%
  \BibitemOpen
  \bibfield  {author} {\bibinfo {author} {\bibfnamefont {B.}~\bibnamefont {Cervantes}}, \bibinfo {author} {\bibfnamefont {G.}~\bibnamefont {Passante}}, \bibinfo {author} {\bibfnamefont {B.~R.}\ \bibnamefont {Wilcox}},\ and\ \bibinfo {author} {\bibfnamefont {S.~J.}\ \bibnamefont {Pollock}},\ }\bibfield  {title} {\bibinfo {title} {An overview of quantum information science courses at {US} institutions},\ }in\ \href@noop {} {\emph {\bibinfo {booktitle} {2021 Physics Education Research Conference Proceedings}}}\ (\bibinfo {year} {2021})\BibitemShut {NoStop}%
\bibitem [{\citenamefont {Buzzell}\ \emph {et~al.}(2025)\citenamefont {Buzzell}, \citenamefont {Atherton},\ and\ \citenamefont {Barthelemy}}]{buzzell2025quantum}%
  \BibitemOpen
  \bibfield  {author} {\bibinfo {author} {\bibfnamefont {A.}~\bibnamefont {Buzzell}}, \bibinfo {author} {\bibfnamefont {T.~J.}\ \bibnamefont {Atherton}},\ and\ \bibinfo {author} {\bibfnamefont {R.}~\bibnamefont {Barthelemy}},\ }\bibfield  {title} {\bibinfo {title} {Quantum mechanics curriculum in the us: Quantifying the instructional time, content taught, and paradigms used},\ }\href@noop {} {\bibfield  {journal} {\bibinfo  {journal} {Physical Review Physics Education Research}\ }\textbf {\bibinfo {volume} {21}},\ \bibinfo {pages} {010102} (\bibinfo {year} {2025})}\BibitemShut {NoStop}%
\bibitem [{\citenamefont {Pi{\~n}a}\ \emph {et~al.}(2025{\natexlab{a}})\citenamefont {Pi{\~n}a}, \citenamefont {El-Adawy}, \citenamefont {Verostek}, \citenamefont {Boyle}, \citenamefont {Cacheiro}, \citenamefont {Lawler}, \citenamefont {Pradeep}, \citenamefont {Watts}, \citenamefont {West}, \citenamefont {Lewandowski} \emph {et~al.}}]{Pina2025}%
  \BibitemOpen
  \bibfield  {author} {\bibinfo {author} {\bibfnamefont {A.}~\bibnamefont {Pi{\~n}a}}, \bibinfo {author} {\bibfnamefont {S.}~\bibnamefont {El-Adawy}}, \bibinfo {author} {\bibfnamefont {M.}~\bibnamefont {Verostek}}, \bibinfo {author} {\bibfnamefont {B.~T.}\ \bibnamefont {Boyle}}, \bibinfo {author} {\bibfnamefont {M.}~\bibnamefont {Cacheiro}}, \bibinfo {author} {\bibfnamefont {M.}~\bibnamefont {Lawler}}, \bibinfo {author} {\bibfnamefont {N.}~\bibnamefont {Pradeep}}, \bibinfo {author} {\bibfnamefont {E.}~\bibnamefont {Watts}}, \bibinfo {author} {\bibfnamefont {C.~G.}\ \bibnamefont {West}}, \bibinfo {author} {\bibfnamefont {H.}~\bibnamefont {Lewandowski}}, \emph {et~al.},\ }\bibfield  {title} {\bibinfo {title} {Landscape of quantum information science and engineering education: From physics foundations to interdisciplinary frontiers},\ }\href@noop {} {\bibfield  {journal} {\bibinfo  {journal} {arXiv preprint arXiv:2504.13719}\ } (\bibinfo {year} {2025}{\natexlab{a}})}\BibitemShut {NoStop}%
\bibitem [{\citenamefont {Fox}\ \emph {et~al.}(2020)\citenamefont {Fox}, \citenamefont {Zwickl},\ and\ \citenamefont {Lewandowski}}]{fox2020preparing}%
  \BibitemOpen
  \bibfield  {author} {\bibinfo {author} {\bibfnamefont {M.~F.}\ \bibnamefont {Fox}}, \bibinfo {author} {\bibfnamefont {B.~M.}\ \bibnamefont {Zwickl}},\ and\ \bibinfo {author} {\bibfnamefont {H.~J.}\ \bibnamefont {Lewandowski}},\ }\bibfield  {title} {\bibinfo {title} {Preparing for the quantum revolution: What is the role of higher education?},\ }\href@noop {} {\bibfield  {journal} {\bibinfo  {journal} {Physical Review Physics Education Research}\ }\textbf {\bibinfo {volume} {16}},\ \bibinfo {pages} {020131} (\bibinfo {year} {2020})}\BibitemShut {NoStop}%
\bibitem [{\citenamefont {Aiello}\ \emph {et~al.}(2021)\citenamefont {Aiello}, \citenamefont {Awschalom}, \citenamefont {Bernien}, \citenamefont {Brower}, \citenamefont {Brown}, \citenamefont {Brun}, \citenamefont {Caram}, \citenamefont {Chitambar}, \citenamefont {Di~Felice}, \citenamefont {Edmonds} \emph {et~al.}}]{aiello2021achieving}%
  \BibitemOpen
  \bibfield  {author} {\bibinfo {author} {\bibfnamefont {C.~D.}\ \bibnamefont {Aiello}}, \bibinfo {author} {\bibfnamefont {D.~D.}\ \bibnamefont {Awschalom}}, \bibinfo {author} {\bibfnamefont {H.}~\bibnamefont {Bernien}}, \bibinfo {author} {\bibfnamefont {T.}~\bibnamefont {Brower}}, \bibinfo {author} {\bibfnamefont {K.~R.}\ \bibnamefont {Brown}}, \bibinfo {author} {\bibfnamefont {T.~A.}\ \bibnamefont {Brun}}, \bibinfo {author} {\bibfnamefont {J.~R.}\ \bibnamefont {Caram}}, \bibinfo {author} {\bibfnamefont {E.}~\bibnamefont {Chitambar}}, \bibinfo {author} {\bibfnamefont {R.}~\bibnamefont {Di~Felice}}, \bibinfo {author} {\bibfnamefont {K.~M.}\ \bibnamefont {Edmonds}}, \emph {et~al.},\ }\bibfield  {title} {\bibinfo {title} {Achieving a quantum smart workforce},\ }\href@noop {} {\bibfield  {journal} {\bibinfo  {journal} {Quantum Science and Technology}\ }\textbf {\bibinfo {volume} {6}},\ \bibinfo {pages} {030501} (\bibinfo {year} {2021})}\BibitemShut {NoStop}%
\bibitem [{\citenamefont {Kaur}\ and\ \citenamefont {Venegas-Gomez}(2022)}]{kaur2022defining}%
  \BibitemOpen
  \bibfield  {author} {\bibinfo {author} {\bibfnamefont {M.}~\bibnamefont {Kaur}}\ and\ \bibinfo {author} {\bibfnamefont {A.}~\bibnamefont {Venegas-Gomez}},\ }\bibfield  {title} {\bibinfo {title} {Defining the quantum workforce landscape: a review of global quantum education initiatives},\ }\href@noop {} {\bibfield  {journal} {\bibinfo  {journal} {Optical Engineering}\ }\textbf {\bibinfo {volume} {61}},\ \bibinfo {pages} {081806} (\bibinfo {year} {2022})}\BibitemShut {NoStop}%
\bibitem [{\citenamefont {Hughes}\ \emph {et~al.}(2022)\citenamefont {Hughes}, \citenamefont {Finke}, \citenamefont {German}, \citenamefont {Merzbacher}, \citenamefont {Vora},\ and\ \citenamefont {Lewandowski}}]{hughes2022assessing}%
  \BibitemOpen
  \bibfield  {author} {\bibinfo {author} {\bibfnamefont {C.}~\bibnamefont {Hughes}}, \bibinfo {author} {\bibfnamefont {D.}~\bibnamefont {Finke}}, \bibinfo {author} {\bibfnamefont {D.-A.}\ \bibnamefont {German}}, \bibinfo {author} {\bibfnamefont {C.}~\bibnamefont {Merzbacher}}, \bibinfo {author} {\bibfnamefont {P.~M.}\ \bibnamefont {Vora}},\ and\ \bibinfo {author} {\bibfnamefont {H.}~\bibnamefont {Lewandowski}},\ }\bibfield  {title} {\bibinfo {title} {Assessing the needs of the quantum industry},\ }\href@noop {} {\bibfield  {journal} {\bibinfo  {journal} {IEEE Transactions on Education}\ }\textbf {\bibinfo {volume} {65}},\ \bibinfo {pages} {592} (\bibinfo {year} {2022})}\BibitemShut {NoStop}%
\bibitem [{\citenamefont {Hasanovic}\ \emph {et~al.}(2022)\citenamefont {Hasanovic}, \citenamefont {Panayiotou}, \citenamefont {Silberman}, \citenamefont {Stimers},\ and\ \citenamefont {Merzbacher}}]{hasanovic2022quantum}%
  \BibitemOpen
  \bibfield  {author} {\bibinfo {author} {\bibfnamefont {M.}~\bibnamefont {Hasanovic}}, \bibinfo {author} {\bibfnamefont {C.}~\bibnamefont {Panayiotou}}, \bibinfo {author} {\bibfnamefont {D.}~\bibnamefont {Silberman}}, \bibinfo {author} {\bibfnamefont {P.}~\bibnamefont {Stimers}},\ and\ \bibinfo {author} {\bibfnamefont {C.}~\bibnamefont {Merzbacher}},\ }\bibfield  {title} {\bibinfo {title} {Quantum technician skills and competencies for the emerging quantum 2.0 industry},\ }\href@noop {} {\bibfield  {journal} {\bibinfo  {journal} {Optical Engineering}\ }\textbf {\bibinfo {volume} {61}},\ \bibinfo {pages} {081803} (\bibinfo {year} {2022})}\BibitemShut {NoStop}%
\bibitem [{\citenamefont {Greinert}\ \emph {et~al.}(2023{\natexlab{a}})\citenamefont {Greinert}, \citenamefont {M{\"u}ller}, \citenamefont {Bitzenbauer}, \citenamefont {Ubben},\ and\ \citenamefont {Weber}}]{greinert2023future}%
  \BibitemOpen
  \bibfield  {author} {\bibinfo {author} {\bibfnamefont {F.}~\bibnamefont {Greinert}}, \bibinfo {author} {\bibfnamefont {R.}~\bibnamefont {M{\"u}ller}}, \bibinfo {author} {\bibfnamefont {P.}~\bibnamefont {Bitzenbauer}}, \bibinfo {author} {\bibfnamefont {M.~S.}\ \bibnamefont {Ubben}},\ and\ \bibinfo {author} {\bibfnamefont {K.-A.}\ \bibnamefont {Weber}},\ }\bibfield  {title} {\bibinfo {title} {Future quantum workforce: Competences, requirements, and forecasts},\ }\href@noop {} {\bibfield  {journal} {\bibinfo  {journal} {Physical Review Physics Education Research}\ }\textbf {\bibinfo {volume} {19}},\ \bibinfo {pages} {010137} (\bibinfo {year} {2023}{\natexlab{a}})}\BibitemShut {NoStop}%
\bibitem [{\citenamefont {Greinert}\ \emph {et~al.}(2023{\natexlab{b}})\citenamefont {Greinert}, \citenamefont {M{\"u}ller}, \citenamefont {Goorney}, \citenamefont {Sherson},\ and\ \citenamefont {Ubben}}]{greinert2023towards}%
  \BibitemOpen
  \bibfield  {author} {\bibinfo {author} {\bibfnamefont {F.}~\bibnamefont {Greinert}}, \bibinfo {author} {\bibfnamefont {R.}~\bibnamefont {M{\"u}ller}}, \bibinfo {author} {\bibfnamefont {S.}~\bibnamefont {Goorney}}, \bibinfo {author} {\bibfnamefont {J.}~\bibnamefont {Sherson}},\ and\ \bibinfo {author} {\bibfnamefont {M.~S.}\ \bibnamefont {Ubben}},\ }\bibfield  {title} {\bibinfo {title} {Towards a quantum ready workforce: the updated european competence framework for quantum technologies},\ }\href@noop {} {\bibfield  {journal} {\bibinfo  {journal} {Frontiers in Quantum Science and Technology}\ }\textbf {\bibinfo {volume} {2}},\ \bibinfo {pages} {1225733} (\bibinfo {year} {2023}{\natexlab{b}})}\BibitemShut {NoStop}%
\bibitem [{\citenamefont {Greinert}\ \emph {et~al.}(2024)\citenamefont {Greinert}, \citenamefont {Ubben}, \citenamefont {Dogan}, \citenamefont {Hilfert-R{\"u}ppell},\ and\ \citenamefont {M{\"u}ller}}]{greinert2024advancing}%
  \BibitemOpen
  \bibfield  {author} {\bibinfo {author} {\bibfnamefont {F.}~\bibnamefont {Greinert}}, \bibinfo {author} {\bibfnamefont {M.~S.}\ \bibnamefont {Ubben}}, \bibinfo {author} {\bibfnamefont {I.~N.}\ \bibnamefont {Dogan}}, \bibinfo {author} {\bibfnamefont {D.}~\bibnamefont {Hilfert-R{\"u}ppell}},\ and\ \bibinfo {author} {\bibfnamefont {R.}~\bibnamefont {M{\"u}ller}},\ }\bibfield  {title} {\bibinfo {title} {Advancing quantum technology workforce: industry insights into qualification and training needs},\ }\href@noop {} {\bibfield  {journal} {\bibinfo  {journal} {EPJ Quantum Technology}\ }\textbf {\bibinfo {volume} {11}},\ \bibinfo {pages} {82} (\bibinfo {year} {2024})}\BibitemShut {NoStop}%
\bibitem [{\citenamefont {Meyer}\ \emph {et~al.}(2024{\natexlab{a}})\citenamefont {Meyer}, \citenamefont {Passante}, \citenamefont {Pollock},\ and\ \citenamefont {Wilcox}}]{meyer2024introductory}%
  \BibitemOpen
  \bibfield  {author} {\bibinfo {author} {\bibfnamefont {J.~C.}\ \bibnamefont {Meyer}}, \bibinfo {author} {\bibfnamefont {G.}~\bibnamefont {Passante}}, \bibinfo {author} {\bibfnamefont {S.~J.}\ \bibnamefont {Pollock}},\ and\ \bibinfo {author} {\bibfnamefont {B.~R.}\ \bibnamefont {Wilcox}},\ }\bibfield  {title} {\bibinfo {title} {Introductory quantum information science coursework at us institutions: content coverage},\ }\href@noop {} {\bibfield  {journal} {\bibinfo  {journal} {EPJ Quantum Technology}\ }\textbf {\bibinfo {volume} {11}},\ \bibinfo {pages} {16} (\bibinfo {year} {2024}{\natexlab{a}})}\BibitemShut {NoStop}%
\bibitem [{\citenamefont {Meyer}\ \emph {et~al.}(2022)\citenamefont {Meyer}, \citenamefont {Passante}, \citenamefont {Pollock},\ and\ \citenamefont {Wilcox}}]{meyer2022today}%
  \BibitemOpen
  \bibfield  {author} {\bibinfo {author} {\bibfnamefont {J.~C.}\ \bibnamefont {Meyer}}, \bibinfo {author} {\bibfnamefont {G.}~\bibnamefont {Passante}}, \bibinfo {author} {\bibfnamefont {S.~J.}\ \bibnamefont {Pollock}},\ and\ \bibinfo {author} {\bibfnamefont {B.~R.}\ \bibnamefont {Wilcox}},\ }\bibfield  {title} {\bibinfo {title} {Today’s interdisciplinary quantum information classroom: Themes from a survey of quantum information science instructors},\ }\href@noop {} {\bibfield  {journal} {\bibinfo  {journal} {Physical Review Physics Education Research}\ }\textbf {\bibinfo {volume} {18}},\ \bibinfo {pages} {010150} (\bibinfo {year} {2022})}\BibitemShut {NoStop}%
\bibitem [{\citenamefont {Barnes}\ \emph {et~al.}(2025)\citenamefont {Barnes}, \citenamefont {Bennett}, \citenamefont {Boltasseva}, \citenamefont {Borish}, \citenamefont {Brown}, \citenamefont {Carr}, \citenamefont {Ceballos}, \citenamefont {Dukes}, \citenamefont {Easton}, \citenamefont {Economou} \emph {et~al.}}]{barnes2025outcomes}%
  \BibitemOpen
  \bibfield  {author} {\bibinfo {author} {\bibfnamefont {E.}~\bibnamefont {Barnes}}, \bibinfo {author} {\bibfnamefont {M.~B.}\ \bibnamefont {Bennett}}, \bibinfo {author} {\bibfnamefont {A.}~\bibnamefont {Boltasseva}}, \bibinfo {author} {\bibfnamefont {V.}~\bibnamefont {Borish}}, \bibinfo {author} {\bibfnamefont {B.}~\bibnamefont {Brown}}, \bibinfo {author} {\bibfnamefont {L.~D.}\ \bibnamefont {Carr}}, \bibinfo {author} {\bibfnamefont {R.~R.}\ \bibnamefont {Ceballos}}, \bibinfo {author} {\bibfnamefont {F.}~\bibnamefont {Dukes}}, \bibinfo {author} {\bibfnamefont {E.~W.}\ \bibnamefont {Easton}}, \bibinfo {author} {\bibfnamefont {S.~E.}\ \bibnamefont {Economou}}, \emph {et~al.},\ }\bibfield  {title} {\bibinfo {title} {Outcomes from a workshop on a national center for quantum education},\ }\href@noop {} {\bibfield  {journal} {\bibinfo  {journal} {EPJ Quantum Technology}\ }\textbf {\bibinfo {volume} {12}},\ \bibinfo {pages} {40} (\bibinfo {year} {2025})}\BibitemShut {NoStop}%
\bibitem [{\citenamefont {Goorney}\ \emph {et~al.}(2024{\natexlab{b}})\citenamefont {Goorney}, \citenamefont {Bley}, \citenamefont {Heusler},\ and\ \citenamefont {Sherson}}]{goorney2024framework}%
  \BibitemOpen
  \bibfield  {author} {\bibinfo {author} {\bibfnamefont {S.}~\bibnamefont {Goorney}}, \bibinfo {author} {\bibfnamefont {J.}~\bibnamefont {Bley}}, \bibinfo {author} {\bibfnamefont {S.}~\bibnamefont {Heusler}},\ and\ \bibinfo {author} {\bibfnamefont {J.}~\bibnamefont {Sherson}},\ }\bibfield  {title} {\bibinfo {title} {A framework for curriculum transformation in quantum information science and technology education},\ }\href@noop {} {\bibfield  {journal} {\bibinfo  {journal} {European Journal of Physics}\ }\textbf {\bibinfo {volume} {45}},\ \bibinfo {pages} {065702} (\bibinfo {year} {2024}{\natexlab{b}})}\BibitemShut {NoStop}%
\bibitem [{\citenamefont {Meyer}\ \emph {et~al.}(2024{\natexlab{b}})\citenamefont {Meyer}, \citenamefont {Passante},\ and\ \citenamefont {Wilcox}}]{meyer2024disparities}%
  \BibitemOpen
  \bibfield  {author} {\bibinfo {author} {\bibfnamefont {J.~C.}\ \bibnamefont {Meyer}}, \bibinfo {author} {\bibfnamefont {G.}~\bibnamefont {Passante}},\ and\ \bibinfo {author} {\bibfnamefont {B.}~\bibnamefont {Wilcox}},\ }\bibfield  {title} {\bibinfo {title} {Disparities in access to us quantum information education},\ }\href@noop {} {\bibfield  {journal} {\bibinfo  {journal} {Physical Review Physics Education Research}\ }\textbf {\bibinfo {volume} {20}},\ \bibinfo {pages} {010131} (\bibinfo {year} {2024}{\natexlab{b}})}\BibitemShut {NoStop}%
\bibitem [{\citenamefont {Morrison}\ \emph {et~al.}(2019)\citenamefont {Morrison}, \citenamefont {Ross}, \citenamefont {Morrison},\ and\ \citenamefont {Kalman}}]{ADDIE}%
  \BibitemOpen
  \bibfield  {author} {\bibinfo {author} {\bibfnamefont {G.~R.}\ \bibnamefont {Morrison}}, \bibinfo {author} {\bibfnamefont {S.~J.}\ \bibnamefont {Ross}}, \bibinfo {author} {\bibfnamefont {J.~R.}\ \bibnamefont {Morrison}},\ and\ \bibinfo {author} {\bibfnamefont {H.~K.}\ \bibnamefont {Kalman}},\ }\href@noop {} {\emph {\bibinfo {title} {Designing effective instruction}}}\ (\bibinfo  {publisher} {John Wiley \& Sons},\ \bibinfo {year} {2019})\BibitemShut {NoStop}%
\bibitem [{\citenamefont {Chasteen}\ \emph {et~al.}(2011)\citenamefont {Chasteen}, \citenamefont {Perkins}, \citenamefont {Beale}, \citenamefont {Pollock},\ and\ \citenamefont {Wieman}}]{chasteen2011thoughtful}%
  \BibitemOpen
  \bibfield  {author} {\bibinfo {author} {\bibfnamefont {S.}~\bibnamefont {Chasteen}}, \bibinfo {author} {\bibfnamefont {K.}~\bibnamefont {Perkins}}, \bibinfo {author} {\bibfnamefont {P.}~\bibnamefont {Beale}}, \bibinfo {author} {\bibfnamefont {S.}~\bibnamefont {Pollock}},\ and\ \bibinfo {author} {\bibfnamefont {C.}~\bibnamefont {Wieman}},\ }\bibfield  {title} {\bibinfo {title} {A thoughtful approach to instruction: Course transformation for the rest of us},\ }\href@noop {} {\bibfield  {journal} {\bibinfo  {journal} {Journal of College Science Teaching}\ } (\bibinfo {year} {2011})}\BibitemShut {NoStop}%
\bibitem [{\citenamefont {Chasteen}\ \emph {et~al.}(2015)\citenamefont {Chasteen}, \citenamefont {Wilcox}, \citenamefont {Caballero}, \citenamefont {Perkins}, \citenamefont {Pollock},\ and\ \citenamefont {Wieman}}]{chasteen2015educational}%
  \BibitemOpen
  \bibfield  {author} {\bibinfo {author} {\bibfnamefont {S.~V.}\ \bibnamefont {Chasteen}}, \bibinfo {author} {\bibfnamefont {B.}~\bibnamefont {Wilcox}}, \bibinfo {author} {\bibfnamefont {M.~D.}\ \bibnamefont {Caballero}}, \bibinfo {author} {\bibfnamefont {K.~K.}\ \bibnamefont {Perkins}}, \bibinfo {author} {\bibfnamefont {S.~J.}\ \bibnamefont {Pollock}},\ and\ \bibinfo {author} {\bibfnamefont {C.~E.}\ \bibnamefont {Wieman}},\ }\bibfield  {title} {\bibinfo {title} {Educational transformation in upper-division physics: The science education initiative model, outcomes, and lessons learned},\ }\href@noop {} {\bibfield  {journal} {\bibinfo  {journal} {Physical Review Special Topics-Physics Education Research}\ }\textbf {\bibinfo {volume} {11}},\ \bibinfo {pages} {020110} (\bibinfo {year} {2015})}\BibitemShut {NoStop}%
\bibitem [{\citenamefont {Reinholz}\ and\ \citenamefont {Apkarian}(2018)}]{reinholz2018four}%
  \BibitemOpen
  \bibfield  {author} {\bibinfo {author} {\bibfnamefont {D.~L.}\ \bibnamefont {Reinholz}}\ and\ \bibinfo {author} {\bibfnamefont {N.}~\bibnamefont {Apkarian}},\ }\bibfield  {title} {\bibinfo {title} {Four frames for systemic change in stem departments},\ }\href@noop {} {\bibfield  {journal} {\bibinfo  {journal} {International Journal of STEM Education}\ }\textbf {\bibinfo {volume} {5}},\ \bibinfo {pages} {1} (\bibinfo {year} {2018})}\BibitemShut {NoStop}%
\bibitem [{\citenamefont {Kania}\ and\ \citenamefont {Kramer}(2011)}]{Kania2011}%
  \BibitemOpen
  \bibfield  {author} {\bibinfo {author} {\bibfnamefont {J.}~\bibnamefont {Kania}}\ and\ \bibinfo {author} {\bibfnamefont {M.}~\bibnamefont {Kramer}},\ }\bibfield  {title} {\bibinfo {title} {Collective impact},\ }\href@noop {} {\bibfield  {journal} {\bibinfo  {journal} {Stanford Social Innovation Review}\ }\textbf {\bibinfo {volume} {9}},\ \bibinfo {pages} {36} (\bibinfo {year} {2011})}\BibitemShut {NoStop}%
\bibitem [{\citenamefont {{ABET}}(2025)}]{ABETAccreditation}%
  \BibitemOpen
  \bibfield  {author} {\bibinfo {author} {\bibnamefont {{ABET}}},\ }\href {https://www.abet.org/accreditation/get-accredited/accreditation-step-by-step/} {\bibinfo {title} {Accreditation step-by-step}} (\bibinfo {year} {2025})\BibitemShut {NoStop}%
\bibitem [{\citenamefont {{HLC}}(2025)}]{HLC}%
  \BibitemOpen
  \bibfield  {author} {\bibinfo {author} {\bibnamefont {{HLC}}},\ }\href {https://www.hlcommission.org} {\bibinfo {title} {Higher learning commission}} (\bibinfo {year} {2025})\BibitemShut {NoStop}%
\bibitem [{\citenamefont {Khanna}\ \emph {et~al.}(2021)\citenamefont {Khanna}, \citenamefont {Roberts},\ and\ \citenamefont {Lane}}]{khanna2021designing}%
  \BibitemOpen
  \bibfield  {author} {\bibinfo {author} {\bibfnamefont {P.}~\bibnamefont {Khanna}}, \bibinfo {author} {\bibfnamefont {C.}~\bibnamefont {Roberts}},\ and\ \bibinfo {author} {\bibfnamefont {A.~S.}\ \bibnamefont {Lane}},\ }\bibfield  {title} {\bibinfo {title} {Designing health professional education curricula using systems thinking perspectives},\ }\href@noop {} {\bibfield  {journal} {\bibinfo  {journal} {BMC Medical Education}\ }\textbf {\bibinfo {volume} {21}},\ \bibinfo {pages} {1} (\bibinfo {year} {2021})}\BibitemShut {NoStop}%
\bibitem [{\citenamefont {Sammut-Bonnici}\ and\ \citenamefont {Galea}(2015)}]{sammut2015swot}%
  \BibitemOpen
  \bibfield  {author} {\bibinfo {author} {\bibfnamefont {T.}~\bibnamefont {Sammut-Bonnici}}\ and\ \bibinfo {author} {\bibfnamefont {D.}~\bibnamefont {Galea}},\ }\bibfield  {title} {\bibinfo {title} {Swot analysis},\ }\href@noop {} {\bibfield  {journal} {\bibinfo  {journal} {Wiley Encyclopedia of management}\ }\textbf {\bibinfo {volume} {12}} (\bibinfo {year} {2015})}\BibitemShut {NoStop}%
\bibitem [{\citenamefont {Orr}(2013)}]{orr2013conducting}%
  \BibitemOpen
  \bibfield  {author} {\bibinfo {author} {\bibfnamefont {B.}~\bibnamefont {Orr}},\ }\bibfield  {title} {\bibinfo {title} {Conducting a swot analysis for program improvement.},\ }\href@noop {} {\bibfield  {journal} {\bibinfo  {journal} {Online Submission}\ }\textbf {\bibinfo {volume} {3}},\ \bibinfo {pages} {381} (\bibinfo {year} {2013})}\BibitemShut {NoStop}%
\bibitem [{\citenamefont {Awuzie}\ \emph {et~al.}(2021)\citenamefont {Awuzie}, \citenamefont {Ngowi}, \citenamefont {Omotayo}, \citenamefont {Obi},\ and\ \citenamefont {Akotia}}]{awuzie2021facilitating}%
  \BibitemOpen
  \bibfield  {author} {\bibinfo {author} {\bibfnamefont {B.}~\bibnamefont {Awuzie}}, \bibinfo {author} {\bibfnamefont {A.~B.}\ \bibnamefont {Ngowi}}, \bibinfo {author} {\bibfnamefont {T.}~\bibnamefont {Omotayo}}, \bibinfo {author} {\bibfnamefont {L.}~\bibnamefont {Obi}},\ and\ \bibinfo {author} {\bibfnamefont {J.}~\bibnamefont {Akotia}},\ }\bibfield  {title} {\bibinfo {title} {Facilitating successful smart campus transitions: A systems thinking-swot analysis approach},\ }\href@noop {} {\bibfield  {journal} {\bibinfo  {journal} {Applied Sciences}\ }\textbf {\bibinfo {volume} {11}},\ \bibinfo {pages} {2044} (\bibinfo {year} {2021})}\BibitemShut {NoStop}%
\bibitem [{\citenamefont {Allareddy}\ \emph {et~al.}(2024)\citenamefont {Allareddy}, \citenamefont {Atsawasuwan}, \citenamefont {Frazier-Bowers}, \citenamefont {Hong}, \citenamefont {Huja}, \citenamefont {Katebi}, \citenamefont {Lee}, \citenamefont {Mehta}, \citenamefont {Padala}, \citenamefont {Utreja} \emph {et~al.}}]{allareddy2024orthodontic}%
  \BibitemOpen
  \bibfield  {author} {\bibinfo {author} {\bibfnamefont {V.}~\bibnamefont {Allareddy}}, \bibinfo {author} {\bibfnamefont {P.}~\bibnamefont {Atsawasuwan}}, \bibinfo {author} {\bibfnamefont {S.}~\bibnamefont {Frazier-Bowers}}, \bibinfo {author} {\bibfnamefont {C.}~\bibnamefont {Hong}}, \bibinfo {author} {\bibfnamefont {S.}~\bibnamefont {Huja}}, \bibinfo {author} {\bibfnamefont {N.}~\bibnamefont {Katebi}}, \bibinfo {author} {\bibfnamefont {M.~K.}\ \bibnamefont {Lee}}, \bibinfo {author} {\bibfnamefont {S.~Y.}\ \bibnamefont {Mehta}}, \bibinfo {author} {\bibfnamefont {S.}~\bibnamefont {Padala}}, \bibinfo {author} {\bibfnamefont {A.}~\bibnamefont {Utreja}}, \emph {et~al.},\ }\bibfield  {title} {\bibinfo {title} {Orthodontic educational landscape in the contemporary context: Insights from educators},\ }in\ \href@noop {} {\emph {\bibinfo {booktitle} {Seminars in Orthodontics}}}\ (\bibinfo {organization} {Elsevier},\ \bibinfo {year} {2024})\BibitemShut {NoStop}%
\bibitem [{\citenamefont {Etikan}\ \emph {et~al.}(2016)\citenamefont {Etikan}, \citenamefont {Musa}, \citenamefont {Alkassim} \emph {et~al.}}]{etikan2016comparison}%
  \BibitemOpen
  \bibfield  {author} {\bibinfo {author} {\bibfnamefont {I.}~\bibnamefont {Etikan}}, \bibinfo {author} {\bibfnamefont {S.~A.}\ \bibnamefont {Musa}}, \bibinfo {author} {\bibfnamefont {R.~S.}\ \bibnamefont {Alkassim}}, \emph {et~al.},\ }\bibfield  {title} {\bibinfo {title} {Comparison of convenience sampling and purposive sampling},\ }\href@noop {} {\bibfield  {journal} {\bibinfo  {journal} {American journal of theoretical and applied statistics}\ }\textbf {\bibinfo {volume} {5}},\ \bibinfo {pages} {1} (\bibinfo {year} {2016})}\BibitemShut {NoStop}%
\bibitem [{\citenamefont {Pi{\~n}a}\ \emph {et~al.}(2025{\natexlab{b}})\citenamefont {Pi{\~n}a}, \citenamefont {El-Adawy}, \citenamefont {Verostek}, \citenamefont {Lewandowski},\ and\ \citenamefont {Zwickl}}]{pina2025asee}%
  \BibitemOpen
  \bibfield  {author} {\bibinfo {author} {\bibfnamefont {A.~R.}\ \bibnamefont {Pi{\~n}a}}, \bibinfo {author} {\bibfnamefont {S.}~\bibnamefont {El-Adawy}}, \bibinfo {author} {\bibfnamefont {M.}~\bibnamefont {Verostek}}, \bibinfo {author} {\bibfnamefont {H.~J.}\ \bibnamefont {Lewandowski}},\ and\ \bibinfo {author} {\bibfnamefont {B.~M.}\ \bibnamefont {Zwickl}},\ }\bibfield  {title} {\bibinfo {title} {Investigating opportunities for growth and increased diversity in quantum information science and engineering education in the {U.S.} based on an analysis of the current educational landscape},\ }\href {https://arxiv.org/abs/2505.00104} {\bibfield  {journal} {\bibinfo  {journal} {arXiv preprint arXiv:2505.00104}\ } (\bibinfo {year} {2025}{\natexlab{b}})},\ \Eprint {https://arxiv.org/abs/2505.00104} {arXiv:2505.00104} \BibitemShut {NoStop}%
\bibitem [{\citenamefont {Zwickl}\ \emph {et~al.}(2024)\citenamefont {Zwickl}, \citenamefont {Howland},\ and\ \citenamefont {Pradeep}}]{Zwickl2024QuantumSensing}%
  \BibitemOpen
  \bibfield  {author} {\bibinfo {author} {\bibfnamefont {B.~M.}\ \bibnamefont {Zwickl}}, \bibinfo {author} {\bibfnamefont {G.~A.}\ \bibnamefont {Howland}},\ and\ \bibinfo {author} {\bibfnamefont {N.}~\bibnamefont {Pradeep}},\ }\bibfield  {title} {\bibinfo {title} {Mapping expert knowledge of quantum sensing and measurement},\ }in\ \href {https://meetings.aps.org/Meeting/APR24/Session/F08.3} {\emph {\bibinfo {booktitle} {Bulletin of the American Physical Society}}}\ (\bibinfo {year} {2024})\BibitemShut {NoStop}%
\bibitem [{\citenamefont {Pradeep}\ \emph {et~al.}(2025)\citenamefont {Pradeep}, \citenamefont {Zwickl},\ and\ \citenamefont {Howland}}]{Namitha2025}%
  \BibitemOpen
  \bibfield  {author} {\bibinfo {author} {\bibfnamefont {N.}~\bibnamefont {Pradeep}}, \bibinfo {author} {\bibfnamefont {B.~M.}\ \bibnamefont {Zwickl}},\ and\ \bibinfo {author} {\bibfnamefont {G.~A.}\ \bibnamefont {Howland}},\ }\bibfield  {title} {\bibinfo {title} {Analysis of undergraduate physics and quantum computing textbooks for concepts related to quantum sensing},\ }in\ \href {https://summit.aps.org/events/APR-S22/2} {\emph {\bibinfo {booktitle} {Bulletin of the American Physical Society}}}\ (\bibinfo {year} {2025})\BibitemShut {NoStop}%
\bibitem [{\citenamefont {McKagan}\ \emph {et~al.}(2020)\citenamefont {McKagan}, \citenamefont {Strubbe}, \citenamefont {Barbato}, \citenamefont {Mason}, \citenamefont {Madsen},\ and\ \citenamefont {Sayre}}]{mckagan2020physport}%
  \BibitemOpen
  \bibfield  {author} {\bibinfo {author} {\bibfnamefont {S.~B.}\ \bibnamefont {McKagan}}, \bibinfo {author} {\bibfnamefont {L.~E.}\ \bibnamefont {Strubbe}}, \bibinfo {author} {\bibfnamefont {L.~J.}\ \bibnamefont {Barbato}}, \bibinfo {author} {\bibfnamefont {B.~A.}\ \bibnamefont {Mason}}, \bibinfo {author} {\bibfnamefont {A.~M.}\ \bibnamefont {Madsen}},\ and\ \bibinfo {author} {\bibfnamefont {E.~C.}\ \bibnamefont {Sayre}},\ }\bibfield  {title} {\bibinfo {title} {Physport use and growth: Supporting physics teaching with research-based resources since 2011},\ }\href@noop {} {\bibfield  {journal} {\bibinfo  {journal} {The Physics Teacher}\ }\textbf {\bibinfo {volume} {58}},\ \bibinfo {pages} {465} (\bibinfo {year} {2020})}\BibitemShut {NoStop}%
\bibitem [{\citenamefont {Chessey}\ \emph {et~al.}(2023)\citenamefont {Chessey}, \citenamefont {Turpen}, \citenamefont {Madsen},\ and\ \citenamefont {McKagan}}]{chessey2023living}%
  \BibitemOpen
  \bibfield  {author} {\bibinfo {author} {\bibfnamefont {M.~K.}\ \bibnamefont {Chessey}}, \bibinfo {author} {\bibfnamefont {C.~A.}\ \bibnamefont {Turpen}}, \bibinfo {author} {\bibfnamefont {A.~M.}\ \bibnamefont {Madsen}},\ and\ \bibinfo {author} {\bibfnamefont {S.~B.}\ \bibnamefont {McKagan}},\ }\bibfield  {title} {\bibinfo {title} {The living physics portal: Reimagining college physics teaching by recognizing faculty expertise and providing opportunities for influence},\ }in\ \href@noop {} {\emph {\bibinfo {booktitle} {AIP Conference Proceedings}}},\ Vol.\ \bibinfo {volume} {3040}\ (\bibinfo {organization} {AIP Publishing},\ \bibinfo {year} {2023})\BibitemShut {NoStop}%
\bibitem [{\citenamefont {Hestenes}\ \emph {et~al.}(1992)\citenamefont {Hestenes}, \citenamefont {Wells}, \citenamefont {Swackhamer} \emph {et~al.}}]{hestenes1992force}%
  \BibitemOpen
  \bibfield  {author} {\bibinfo {author} {\bibfnamefont {D.}~\bibnamefont {Hestenes}}, \bibinfo {author} {\bibfnamefont {M.}~\bibnamefont {Wells}}, \bibinfo {author} {\bibfnamefont {G.}~\bibnamefont {Swackhamer}}, \emph {et~al.},\ }\bibfield  {title} {\bibinfo {title} {Force concept inventory},\ }\href@noop {} {\bibfield  {journal} {\bibinfo  {journal} {The physics teacher}\ }\textbf {\bibinfo {volume} {30}},\ \bibinfo {pages} {141} (\bibinfo {year} {1992})}\BibitemShut {NoStop}%
\bibitem [{\citenamefont {Madsen}\ \emph {et~al.}(2017)\citenamefont {Madsen}, \citenamefont {McKagan},\ and\ \citenamefont {Sayre}}]{Madsen2017RBAI1}%
  \BibitemOpen
  \bibfield  {author} {\bibinfo {author} {\bibfnamefont {A.~M.}\ \bibnamefont {Madsen}}, \bibinfo {author} {\bibfnamefont {S.~B.}\ \bibnamefont {McKagan}},\ and\ \bibinfo {author} {\bibfnamefont {E.~C.}\ \bibnamefont {Sayre}},\ }\bibfield  {title} {\bibinfo {title} {Resource letter {RBAI‑1}: Research‑based assessment instruments in physics and astronomy},\ }\href {https://doi.org/10.1119/1.4985045} {\bibfield  {journal} {\bibinfo  {journal} {American Journal of Physics}\ }\textbf {\bibinfo {volume} {85}},\ \bibinfo {pages} {245} (\bibinfo {year} {2017})}\BibitemShut {NoStop}%
\bibitem [{\citenamefont {Madsen}\ \emph {et~al.}(2019)\citenamefont {Madsen}, \citenamefont {McKagan},\ and\ \citenamefont {Sayre}}]{Madsen2019RBAI2}%
  \BibitemOpen
  \bibfield  {author} {\bibinfo {author} {\bibfnamefont {A.~M.}\ \bibnamefont {Madsen}}, \bibinfo {author} {\bibfnamefont {S.~B.}\ \bibnamefont {McKagan}},\ and\ \bibinfo {author} {\bibfnamefont {E.~C.}\ \bibnamefont {Sayre}},\ }\bibfield  {title} {\bibinfo {title} {Resource letter {RBAI‑2}: Research‑based assessment instruments: Beyond physics topics},\ }\href {https://doi.org/10.1119/1.5094139} {\bibfield  {journal} {\bibinfo  {journal} {American Journal of Physics}\ }\textbf {\bibinfo {volume} {87}},\ \bibinfo {pages} {350} (\bibinfo {year} {2019})}\BibitemShut {NoStop}%
\bibitem [{\citenamefont {Cataloglu}\ and\ \citenamefont {Robinett}(2002)}]{cataloglu2002testing}%
  \BibitemOpen
  \bibfield  {author} {\bibinfo {author} {\bibfnamefont {E.}~\bibnamefont {Cataloglu}}\ and\ \bibinfo {author} {\bibfnamefont {R.}~\bibnamefont {Robinett}},\ }\bibfield  {title} {\bibinfo {title} {Testing the development of student conceptual and visualization understanding in quantum mechanics through the undergraduate career},\ }\href@noop {} {\bibfield  {journal} {\bibinfo  {journal} {American Journal of Physics}\ }\textbf {\bibinfo {volume} {70}},\ \bibinfo {pages} {238} (\bibinfo {year} {2002})}\BibitemShut {NoStop}%
\bibitem [{\citenamefont {Wuttiprom}\ \emph {et~al.}(2009)\citenamefont {Wuttiprom}, \citenamefont {Sharma}, \citenamefont {Johnston}, \citenamefont {Chitaree},\ and\ \citenamefont {Soankwan}}]{wuttiprom2009development}%
  \BibitemOpen
  \bibfield  {author} {\bibinfo {author} {\bibfnamefont {S.}~\bibnamefont {Wuttiprom}}, \bibinfo {author} {\bibfnamefont {M.~D.}\ \bibnamefont {Sharma}}, \bibinfo {author} {\bibfnamefont {I.~D.}\ \bibnamefont {Johnston}}, \bibinfo {author} {\bibfnamefont {R.}~\bibnamefont {Chitaree}},\ and\ \bibinfo {author} {\bibfnamefont {C.}~\bibnamefont {Soankwan}},\ }\bibfield  {title} {\bibinfo {title} {Development and use of a conceptual survey in introductory quantum physics},\ }\href@noop {} {\bibfield  {journal} {\bibinfo  {journal} {International Journal of Science Education}\ }\textbf {\bibinfo {volume} {31}},\ \bibinfo {pages} {631} (\bibinfo {year} {2009})}\BibitemShut {NoStop}%
\bibitem [{\citenamefont {McKagan}\ \emph {et~al.}(2010)\citenamefont {McKagan}, \citenamefont {Perkins},\ and\ \citenamefont {Wieman}}]{mckagan2010design}%
  \BibitemOpen
  \bibfield  {author} {\bibinfo {author} {\bibfnamefont {S.}~\bibnamefont {McKagan}}, \bibinfo {author} {\bibfnamefont {K.}~\bibnamefont {Perkins}},\ and\ \bibinfo {author} {\bibfnamefont {C.}~\bibnamefont {Wieman}},\ }\bibfield  {title} {\bibinfo {title} {Design and validation of the quantum mechanics conceptual survey},\ }\href@noop {} {\bibfield  {journal} {\bibinfo  {journal} {Physical Review Special Topics—Physics Education Research}\ }\textbf {\bibinfo {volume} {6}},\ \bibinfo {pages} {020121} (\bibinfo {year} {2010})}\BibitemShut {NoStop}%
\bibitem [{\citenamefont {Sadaghiani}\ and\ \citenamefont {Pollock}(2015)}]{sadaghiani2015quantum}%
  \BibitemOpen
  \bibfield  {author} {\bibinfo {author} {\bibfnamefont {H.~R.}\ \bibnamefont {Sadaghiani}}\ and\ \bibinfo {author} {\bibfnamefont {S.~J.}\ \bibnamefont {Pollock}},\ }\bibfield  {title} {\bibinfo {title} {Quantum mechanics concept assessment: Development and validation study},\ }\href@noop {} {\bibfield  {journal} {\bibinfo  {journal} {Physical Review Special Topics-Physics Education Research}\ }\textbf {\bibinfo {volume} {11}},\ \bibinfo {pages} {010110} (\bibinfo {year} {2015})}\BibitemShut {NoStop}%
\bibitem [{\citenamefont {Marshman}\ and\ \citenamefont {Singh}(2019)}]{marshman2019validation}%
  \BibitemOpen
  \bibfield  {author} {\bibinfo {author} {\bibfnamefont {E.}~\bibnamefont {Marshman}}\ and\ \bibinfo {author} {\bibfnamefont {C.}~\bibnamefont {Singh}},\ }\bibfield  {title} {\bibinfo {title} {Validation and administration of a conceptual survey on the formalism and postulates of quantum mechanics},\ }\href@noop {} {\bibfield  {journal} {\bibinfo  {journal} {Physical Review Physics Education Research}\ }\textbf {\bibinfo {volume} {15}},\ \bibinfo {pages} {020128} (\bibinfo {year} {2019})}\BibitemShut {NoStop}%
\bibitem [{\citenamefont {Durkin}\ \emph {et~al.}(2025)\citenamefont {Durkin}, \citenamefont {Lin}, \citenamefont {Kolodrubetz},\ and\ \citenamefont {McMahan}}]{Durkin2025QISCIT}%
  \BibitemOpen
  \bibfield  {author} {\bibinfo {author} {\bibfnamefont {K.}~\bibnamefont {Durkin}}, \bibinfo {author} {\bibfnamefont {M.}~\bibnamefont {Lin}}, \bibinfo {author} {\bibfnamefont {M.~H.}\ \bibnamefont {Kolodrubetz}},\ and\ \bibinfo {author} {\bibfnamefont {R.~P.}\ \bibnamefont {McMahan}},\ }\bibfield  {title} {\bibinfo {title} {Qiscit: A validated concept inventory assessment for quantum information science},\ }\href@noop {} {\bibfield  {journal} {\bibinfo  {journal} {arXiv preprint arXiv:2506.17122}\ } (\bibinfo {year} {2025})},\ \bibinfo {note} {preprint. \\url{https://arxiv.org/abs/2506.17122}}\BibitemShut {NoStop}%
\bibitem [{\citenamefont {{Effective Practices for Physics Programs (EP3)}}(2024)}]{ep3_assess_learning}%
  \BibitemOpen
  \bibfield  {author} {\bibinfo {author} {\bibnamefont {{Effective Practices for Physics Programs (EP3)}}},\ }\href {https://ep3guide.org/guide/how-to-assess-student-learning} {\bibinfo {title} {How to assess student learning}} (\bibinfo {year} {2024})\BibitemShut {NoStop}%
\bibitem [{\citenamefont {{Rochester Institute of Technology and University of Colorado Boulder: United States {QISE} workforce and education landscape project}}(2025)}]{RIT_Industry_Perspectives}%
  \BibitemOpen
  \bibfield  {author} {\bibinfo {author} {\bibnamefont {{Rochester Institute of Technology and University of Colorado Boulder: United States {QISE} workforce and education landscape project}}},\ }\href@noop {} {\bibinfo {title} {Industry perspectives}},\ \bibinfo {howpublished} {\url{https://www.rit.edu/quantumeducationandworkforce/industry-perspectives}} (\bibinfo {year} {2025})\BibitemShut {NoStop}%
\end{thebibliography}%

\end{document}